\documentclass[twocolumn,showpacs,preprintnumbers,superscriptaddress,amsmath,amssymb,epsfig,nofootinbib]{revtex4}

\usepackage{graphicx}
\usepackage{dcolumn}
\usepackage{bm}
\usepackage{latexsym}
\usepackage{longtable}
\usepackage{enumerate}
\usepackage{subfloat}
\usepackage{array}
\usepackage{xcolor}
\usepackage{footmisc}

\renewcommand *\arraystretch{1.5}

\begin{document}


\title{\textbf{An investigation of the tungsten isotopes via thermal neutron capture}}

\author{A.~M.~Hurst}
\email{AMHurst@lbl.gov}
\affiliation{Lawrence Berkeley National Laboratory, Berkeley, California 94720, USA}

\author{R.~B.~Firestone}
\affiliation{Lawrence Berkeley National Laboratory, Berkeley, California 94720, USA}

\author{B.~W.~Sleaford}
\affiliation{Lawrence Livermore National Laboratory, Livermore, California 94550, USA}

\author{N.~C.~Summers}
\affiliation{Lawrence Livermore National Laboratory, Livermore, California 94550, USA}

\author{Zs.~R{\'e}vay}
\affiliation{Centre for Energy Research, Hungarian Academy of Sciences, H-1525 Budapest, Hungary}
\affiliation{Technische Universit{\"a}t M{\"u}nchen, Forschungsneutronenquelle Heinz Maier-Leibnitz (FRM II), Garching, Germany}

\author{L.~Szentmikl{\'o}si}
\affiliation{Centre for Energy Research, Hungarian Academy of Sciences, H-1525 Budapest, Hungary}

\author{M.~S.~Basunia}
\affiliation{Lawrence Berkeley National Laboratory, Berkeley, California 94720, USA}

\author{T.~Belgya}
\affiliation{Centre for Energy Research, Hungarian Academy of Sciences, H-1525 Budapest, Hungary}

\author{J.~E.~Escher}
\affiliation{Lawrence Livermore National Laboratory, Livermore, California 94550, USA}

\author{M.~Krti{\v c}ka}
\affiliation{Charles University in Prague, Faculty of Mathematics and Physics, CZ-180 00 Prague, Czech Republic}


\date{\today}

\begin{abstract}

Total radiative thermal neutron-capture $\gamma$-ray cross sections for the $^{182,183,184,186}$W isotopes were measured using guided neutron beams from the Budapest Research Reactor to induce prompt and delayed $\gamma$ rays from elemental and isotopically-enriched tungsten targets.  These cross sections were determined from the sum of measured $\gamma$-ray cross sections feeding the ground state from low-lying levels below a cutoff energy, E$_{\rm crit}$, where the level scheme is completely known, and continuum $\gamma$ rays from levels above E$_{\rm crit}$, calculated using the Monte Carlo statistical-decay code DICEBOX.  The new cross sections determined in this work for the tungsten nuclides are: $\sigma_{0}(^{182}{\rm W}) = 20.5(14)$~b and $\sigma_{11/2^{+}}(^{183}{\rm W}^{m}, 5.2~{\rm s}) = 0.177(18)$~b; $\sigma_{0}(^{183}{\rm W}) = 9.37(38)$~b and $\sigma_{5^{-}}(^{184}{\rm W}^{m}, 8.33~\mu{\rm s}) = 0.0247(55)$~b; $\sigma_{0}(^{184}{\rm W}) = 1.43(10)$~b and $\sigma_{11/2^{+}}(^{185}{\rm W}^{m}, 1.67~{\rm min}) = 0.0062(16)$~b; and, $\sigma_{0}(^{186}{\rm W}) = 33.33(62)$~b and $\sigma_{9/2^{+}}(^{187}{\rm W}^{m}, 1.38~\mu{\rm s}) = 0.400(16)$~b.  These results are consistent with earlier measurements in the literature.  The $^{186}$W cross section was also independently confirmed from an activation measurement, following the decay of $^{187}$W, yielding values for $\sigma_{0}(^{186}{\rm W})$ that are consistent with  our prompt $\gamma$-ray measurement.  The cross-section measurements were found to be insensitive to choice of level density or photon strength model, and only weakly dependent on E$_{\rm crit}$.  Total radiative-capture widths calculated with DICEBOX showed much greater model dependence, however, the recommended values could be reproduced with selected model choices.  The decay schemes for all tungsten isotopes were improved in these analyses.  We were also able to determine new neutron separation energies from our primary $\gamma$-ray measurements for the respective (n,$\gamma$) compounds: $^{183}$W ($S_{\rm n} = 6190.88(6)$~keV); $^{184}$W ($S_{\rm n} = 7411.11(13)$~keV); $^{185}$W ($S_{\rm n} = 5753.74(5)$~keV); and, $^{187}$W ($S_{\rm n} = 5466.62(7)$~keV).

\end{abstract}


\maketitle

\section{\label{sec:level1}Introduction\protect\\}

Neutron-capture decay-scheme data from the Reference Input Parameter Library (RIPL) \cite{capote:09} are required for nuclear-reaction calculations that are used to generate the Evaluated Nuclear Data File (ENDF)~\cite{chadwick:11}.  These data play a valuable role for both nuclear applications and basic research into the statistical properties of the nucleus including level densities and photon strengths.  They also provide a wealth of structural information including discrete level spins and parities $J^{\pi}$ and $\gamma$-ray branching ratios.  In addition, information on neutron-capture cross sections may also be obtained.  Preliminary capture $\gamma$-ray cross sections were previously measured on natural elemental targets and published in the Evaluated Gamma-ray Activation File (EGAF) \cite{firestone:06}.  For many elements only data for the isotopes with the largest cross sections and/or abundances could be obtained with natural targets.  This paper describes a new campaign to improve the EGAF database by measuring isotopically-enriched targets.

Traditional methods for determining the total radiative thermal neutron-capture cross section, $\sigma_0$, include neutron-transmission and pile-oscillator measurements, both of which require precise knowledge of the neutron flux, and activation measurements which require an accurate decay-scheme normalization.  Large corrections due to epithermal (1~eV to 10~keV), fast ($\gtrsim 10$~keV), and high-energy neutrons ($\gtrsim 1$~MeV) are typically necessary to determine thermal-capture cross sections.  In this work we apply a newer method to determine the total radiative thermal neutron-capture cross sections for the tungsten isotopes using partial thermal neutron-capture $\gamma$-ray cross sections, $\sigma_{\gamma}$, measured with a guided thermal-neutron beam, combined with statistical-model calculations to account for unresolved continuum $\gamma$-rays, as described previously for the palladium \cite{krticka:08}, potassium \cite{firestone:13}, and gadolinium isotopes \cite{choi:13}.  The prompt neutron-capture $\gamma$-rays were measured using both isotopically-enriched $^{182,183,186}$W targets and a natural elemental sample, to determine neutron-capture decay schemes for the compound tungsten nuclides $^{183,184,185,187}$W.  This information was then used to normalize Monte Carlo simulations for the corresponding neutron-capture decay schemes calculated with the statistical-decay code {\small DICEBOX} \cite{becvar:98}.   The neutron-capture $\gamma$-ray cross sections directly populating the ground state (GS) from low-lying levels were summed with the smaller, calculated, quasi-continuum contribution feeding the GS from higher levels to determine $\sigma_{0}$ for each tungsten isotope.  Comparison of the simulated and experimental neutron-capture $\gamma$-ray cross sections populating and depopulating each excited state was also used to improve the tungsten decay schemes with the augmentation of more-complete data: determination of accurate $\gamma$-ray branchings, assessment of multipolarity and $\gamma$-ray mixing ratios ($\delta_{\gamma}$), placements of new $\gamma$-ray transitions, resolution of ambiguous (or tentative) energy-level and $J^{\pi}$ assignments, and neutron-separation energies ($S_{\rm n}$) determined from the observed primary $\gamma$-ray data for $^{183,184,185,187}$W.  Also, as a validation of the current approach, the $\gamma$-decay emission probabilities, $P_{\gamma}$, were determined from the activation $\gamma$-ray cross sections corresponding to $^{187}$W $\beta^{-}$ decay.  These measurements were found to be consistent with the adopted values, reported in the Evaluated Nuclear Structure Data File (ENSDF) \cite{ensdf}, that are based on the work of Marnada \emph{et al}. \cite{marnada:99}.

\begin{table}[b]
\caption{\label{tab:isocomp} Isotopic composition of natural \cite{berglund:11} and enriched tungsten samples used in this work.  The left-most column refers to the principal-enriched component in the sample.}
\begin{tabular}{|c|c|c|c|c|c|}\hline\hline
Sample & Mass [mg] & $^{182}$W [\%] & $^{183}$W [\%] & $^{184}$W [\%] & $^{186}$W [\%] \\ \hline
$^{\rm nat}$W & 240 & 26.50(16) & 14.31(4) & 30.64(2) & 28.43(19) \\
$^{182}$W &274& 92.7(9) & 2.0(3) & 4.8(9) & 0.5(1) \\
$^{183}$W & 180 & 9.0(8) & 74.9(3) & 13.7(5) & 2.4(3) \\
$^{186}$W & 169 & 0.35(3) & $\sim 0$ & $\sim 0$ & 99.65(3) \\ \hline\hline
\end{tabular}
\end{table}

\section{\label{sec:level2}Experiment and Data Analysis\protect\\}

Isotopically-enriched stable and natural tungsten targets were irradiated with a supermirror-guided {\sl near-thermal} neutron beam ($T \sim 120$~K; $E_{\rm beam} \sim 4.2$~meV) at the 10-MW Budapest Research Reactor \cite{rosta:97,rosta:02}.  The isotopic compositions of the enriched samples are shown in Table~\ref{tab:isocomp} and were determined by comparison with the ratios of peak intensities of strong, well-resolved transitions from the different tungsten isotopes in an elemental sample after accounting for their natural abundances.  All enriched samples were oxide powders (WO$_{2}$) that were suspended in the evacuated neutron beam line in Teflon bags.  During bombardment the thermal-neutron flux at the Prompt Gamma Activation Analysis (PGAA) target station was approximately $2.3 \times 10^{6}$~${\rm n} \cdot {\rm cm}^{-2} \cdot {\rm s}^{-1}$.  The PGAA facility is located $\sim 35$~m from the reactor wall in a low-background environment.  The observed deexcitation $\gamma$ rays from the $^{A}$W(n,$\gamma$)$^{A+1}$W reactions were recorded in a single Compton-suppressed $n$-type high-purity germanium (HPGe) detector with a closed-end coaxial-type geometry, positioned $\sim 23.5$~cm from the target location.  The PGAA facility is described in detail in Refs.~\cite{revay:04,szentmiklosi:10}.  Energy and counting-efficiency calibrations of the HPGe detector were accomplished using standard radioactive and reaction sources covering an energy range from approximately $0.05-11$~MeV .  The non-linearity and efficiency curves were generated using the $\gamma$-ray spectroscopy software package {\small HYPERMET-PC} \cite{hypermet}, which was also used to perform peak-fitting analysis of the complex capture-$\gamma$ spectra.

Singles $\gamma$-ray data were collected in these (n,$\gamma$) measurements and peak areas for unresolved doublets, and higher-order multiplets, were divided based on branching ratios reported in the ENSDF \cite{ensdf}.  Internal conversion coefficients for all transitions were calculated with the {\small BRICC} calculator, which is based on the Band Raman prescription \cite{kibedi:08}.

\subsection{\label{sec:standard}Standardization Procedure\protect\\}

Partial neutron-capture $\gamma$-ray cross sections were derived from the measured peak intensities of the tungsten capture-$\gamma$ lines using an internal-standardization procedure where the observed $\gamma$-ray intensities are normalized by scaling to well-known comparator lines \cite{revay:03}.  Here we used tungstic acid (H$_{2}$WO$_{4}$) for standardization \cite{szentmiklosi:pc} where hydrogen was used as the comparator with $\sigma_{\gamma}(2223~{\rm keV})=0.3326(7)$~b \cite{revay:03} with a stoichiometric $2:1$ H to W atomic ratio.  The cross sections of the standardized tungsten transitions are listed in Table~\ref{tab:tungstic}.  Cross sections for the more intense tungsten $\gamma$-ray transitions were measured with a natural elemental WO$_{2}$ target and then normalized to the standardized, strong, well-resolved cross sections from the standardization measurement using the well-known natural abundances \cite{berglund:11}.  Weaker $\gamma$-ray transitions were measured in irradiations of enriched targets and similarly standardized.  Since the tungsten isotopes and the calibration standard cross sections have a pure \textit{1/v} dependence near thermal neutron energies i.e. increasing cross section with lower incident-neutron energy, no correction was necessary for the neutron-beam temperature.

\begin{table}[t]
\caption{\label{tab:tungstic} Elemental cross sections corresponding to strong lines observed in the tungsten compounds following an internal-standardization (n,$\gamma$) measurement with H$_{2}$WO$_{4}$ \cite{szentmiklosi:pc} comprising natural elemental tungsten.}
\begin{tabular}{|c|c|c|}\hline\hline
Compound & $E_{\gamma}$ [keV] & $\sigma_{\gamma}$ [b]\\ \hline
$^{187}$W & 77.39(3) &  0.234(4) \\ 
$^{187}$W & 145.79(3) &  1.344(13) \\ 
$^{187}$W & 273.10(5) &  0.380(4) \\ 
$^{187}$W & 5261.68(6) &  0.653(9) \\ 
$^{183}$W & 6190.78(3) &  0.726(10) \\ \hline\hline
\end{tabular}
\end{table}

\subsection{\label{sec:thickness}Determination of the Effective Thickness\protect\\}

Since the WO$_{2}$ powders used in these measurements have a density of 10.8 g/cm$^{3}$, the intensity of low energy $\gamma$-rays must be corrected for self attenuation within the sample.  To make this correction it is necessary to determine the {\sl effective} sample thickness and calculate the intensity-attenuation coefficients as a function of $\gamma$-ray energy based on the prescription outlined in Ref.~\cite{hubbel:95} using data from {\small XMUDAT} \cite{xmudat}.  For irregular-shaped targets with non-uniform surfaces, such as the oxide powders used here, it is difficult to measure the sample thickness directly.  Thus, to determine the effective WO$_2$ target thicknesses we compared the thin, lower-density (5.6 g/cm$^{3}$), attenuation-corrected tungstic acid target standardization-cross-section data, listed in Table~\ref{tab:tungstic}, to the attenuated cross sections of these same transitions in the WO$_{2}$ targets.  We then iteratively varied the sample thickness of the WO$_{2}$ targets until the calculated attenuation converged with the observed values for all transitions.  An attenuation correction was then applied to all $\gamma$-rays in the spectrum.

\section{\label{sec:level3}Statistical Model Calculations\protect\\}

The Monte Carlo statistical-decay code {\small DICEBOX} \cite{becvar:98} was used to simulate the thermal neutron-capture $\gamma$-ray cascade.  {\small DICEBOX} assumes a generalization of the extreme statistical model, proposed by Bohr \cite{bohr:37} in the description of compound-nucleus formation and its subsequent decay.  In thermal neutron capture the compound nucleus is formed with an excitation energy slightly above the neutron-separation energy threshold where particle evaporation is negligible.  Within this theoretical framework, the {\small DICEBOX} calculation is constrained by the experimental decay scheme known up to a cut-off energy referred to as the critical energy, $E_{\rm crit}$, where all energies, spins and parities, and $\gamma$-ray deexcitations of the levels are regarded as complete and accurate.  The code generates a random set of levels between $E_{\rm crit}$ and the neutron-separation energy according to an {\it a priori} assumed level density (LD) model $\rho(E,J^{\pi})$.  Transitions to and from the quasi continuum to low-lying levels are then determined according to a choice of an {\it a priori} assumed photon strength function (PSF), $f^{(XL)}$, where $XL$ denotes the multipolarity of the transition.  Selection rules are used to determine allowed transitions between all possible permutations of pairs of initial ($E_{i}$) and final ($E_{f}$) states given by $E_{\gamma} = E_{i} - E_{f}$.  The partial radiation widths, $\Gamma_{if}^{XL}$, of the corresponding transition probabilities for non-forbidden transitions are assumed to follow a Porter-Thomas distribution \cite{porter:56}, centered on a mean value according to the expression
\begin{equation}
\langle \Gamma_{if}^{(XL)} \rangle = \frac{f^{(XL)}(E_{\gamma}) E_{\gamma}^{2L+1}}{\rho(E_{i}, J_{i}^{\pi_{i}})}.
\label{eq:2}
\end{equation}
Internal conversion is accounted for using {\small BRICC} \cite{kibedi:08}.  The corresponding simulated decay schemes are called {\sl nuclear realizations}.  Statistical fluctuations in the Porter-Thomas distributions are reflected in the variations between nuclear realizations and provide the uncertainty in the simulation inherent in the Porter-Thomas assumption.  In these calculations we performed 50 separate nuclear realizations, with each realization comprising 100,000 capture-state $\gamma$-ray cascades.

The experimental $\gamma$-ray cross sections depopulating the low-lying levels below $E_{\rm crit}$, can then be used to renormalize the simulated population per neutron capture, from {\small DICEBOX}, to absolute cross sections feeding these levels.  The total radiative thermal neutron-capture cross section $\sigma_{0}$ is determined as
\begin{equation}
\sigma_{0}=\sum \sigma_{\gamma}^{\rm exp}({\rm GS})+\sum \sigma_{\gamma}^{\rm sim}({\rm GS}) = \frac{\sum \sigma_{\gamma}^{\rm exp}({\rm GS})}{1-P({\rm GS})},
\label{eq:3}
\end{equation}
where $\sum \sigma_{\gamma}^{\rm exp}({\rm GS})$ represents the sum of experimental $\gamma$-ray cross sections feeding the ground state in direct single-step transitions, either via a primary GS transition or secondary transition from a level below $E_{\rm crit}$.  The simulated contribution from the quasi continuum above $E_{\rm crit}$ feeding the ground state, $\sum \sigma_{\gamma}^{\rm sim}({\rm GS})$, may also be written as the product of $\sigma_{0}$ and the simulated ground-state population per neutron capture, $P({\rm GS})$, given by {\small DICEBOX} as shown in Equation~\ref{eq:3}.

\section{\label{sec:level3.A}Adopted Models\protect\\}

\begin{table*}[t]
\caption{\label{tab:2} Level density parameters for the CTF ($T$ and $E_{0}$) and BSFG ($a$ and $E_{1}$), pairing energies ($\Delta$), and average resonance spacings ($D_{0}$) used in the tungsten simulations with {\small DICEBOX}, taken from Ref.~\cite{vonegidy:05}.  Mean values of the parameters were used in these calculations as their uncertainties have negligible effect on the result.  See the text for details.}
\begin{tabular}{|c|c|c|c|c|c|c|}
\hline\hline
Compound& $T$ [MeV] & $E_{0}$ [MeV] & $a$ [${\rm MeV}^{-1}$] & $E_{1}$ [MeV] & $\Delta$ [MeV] & $D_{0}$ [eV] \\

\hline
$^{183}$W  & 0.55(2) & $-0.92(17)$ & 19.22(30) & $-0.24(10)$  & 0 & 59.9(61)  \\
$^{184}$W  & 0.58(2) & $-0.64(21)$ & 18.76(30) & 0.08(14) & 0.763 & 12.0(10)  \\
$^{185}$W  & 0.56(1) & $-1.30(14)$ & 19.45(28) & $-0.50(8)$   & 0 & 69.9(69)  \\
$^{187}$W  & 0.57(2) & $-1.63(22)$ & 19.14(36) & $-0.81(13)$  & 0 & 84.8(79)  \\
\hline\hline
\end{tabular}
\end{table*}

The simulated population of the levels below $E_{\rm crit}$ depends upon the assumed experimental decay scheme, the capture-state spin composition, $J = 1/2^{+}$ for even-even targets and $J = J_{\rm gs}(\rm target)\pm 1/2$ for odd-odd and odd-$A$ targets, and the choice of adopted phenomenological LD and PSF models.

\subsection{\label{sec:levelLD}Level Densities\protect\\}

The constant temperature formula (CTF) \cite{gilbert:65} and the back-shifted Fermi gas (BSFG) \cite{newton:56,gilbert:65} models were considered in this work.  Both models embody a statistical procedure describing the increasing cumulative number density of levels $N(E)$ with increasing excitation energy such that,
\begin{equation}
N(E) = \int \rho(E) d(E),
\label{eq:ld_insert1}
\end{equation}
where $\rho(E)$ represents the level density at an excitation energy $E$.  In the CTF model, a constant temperature is assumed over the entire range of nuclear excitation energy that may be explicitly stated as
\begin{equation}
\rho(E,J) = \frac{f(J)}{T} \exp{\left( \frac{E-E_{0}}{T} \right)}.
\label{eq:ld_insert2}
\end{equation}
The nuclear temperature $T$ may be interpreted as the critical temperature necessary for breaking nucleon pairs.  The energy backshift related to proton- and neutron-pairing energies is given by $E_{0}$.  The temperature and backshift-energy parametrizations used in this work are taken from von Egidy and Bucurescu \cite{vonegidy:05} and listed in Table~\ref{tab:2}.  A spin-distribution factor $f(J)$ \cite{gilbert:65} is introduced in Equation~\ref{eq:ld_insert2} and assumed to have the separable form of Ref.~\cite{gilbert:65}
\begin{equation}
f(J) = \frac{2J+1}{2 \sigma_{c}^{2}} \exp \left( - \frac{(J+1/2)^{2}}{2 \sigma_{c}^{2}} \right),
\label{eq:5}
\end{equation}
where $\sigma_{c} = 0.98 \cdot A^{0.29}$ denotes the spin cut-off factor \cite{vonegidy:88}.

The BSFG level density model is based on the assumption that the nucleus behaves like a fluid of fermions and may be written as
\begin{equation}
\rho(E,J) = f(J) \frac{ \exp (2\sqrt{a(E-E_{1})})}{12\sqrt{2}\sigma_{c}a^{1/4}(E-E_{1})^{5/4} }.
\label{eq:4}
\end{equation}
Here, the spin cut-off factor $\sigma_{c}$ is defined with an energy dependence given by
\begin{equation}
\sigma_{c}^{2} = 0.0146 \cdot A^{5/3} \cdot \frac{ 1+\sqrt{1+4a(E-E_{1})} }{2a}.
\label{eq:6}
\end{equation}
Since fermions exhibit a tendency to form pairs, the extra amount of energy required to separate them is accounted for by the introduction of the level density parameter, $E_{1}$, in Equation~\ref{eq:4}, above.  This parameter corresponds to the back-shift in excitation energy, while $a$ represents the shell-model level density parameter that varies approximately with $0.21\cdot A^{0.87}$~MeV$^{-1}$ \cite{dobaczewski:01}.  As with the CTF, the adopted BSFG parameters used in this work have also been taken from von Egidy and Bucurescu~\cite{vonegidy:05} and are presented in Table~\ref{tab:2}.  In that work, the level density parameters were treated as adjustable and determined by fitting the functional forms of Equations~\ref{eq:ld_insert2} and \ref{eq:4}, above, to experimentally-observed neutron resonance spacings in the region of the capture state above the neutron-separation energy.

\subsection{\label{sec:levelPSF}Photon Strength Functions\protect\\}

\begin{figure}[b]
\includegraphics[angle=-90,width=\linewidth]{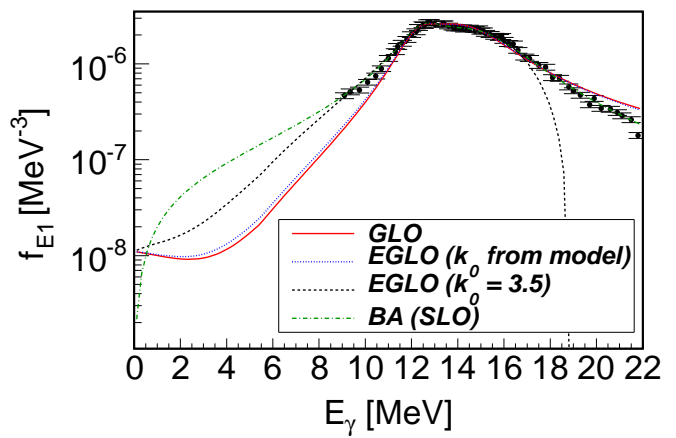}
\caption{\label{fig:2} (Color online) Experimental photon strength functions (PSF) derived from a photoabsorption measurement corresponding to $^{186}$W($\gamma$,abs)~\cite{berman:69}.  The calculated curves represent different theoretical models of the PSF based on the BA \cite{brink:55,axel:62}, GLO \cite{kopecky:90}, and EGLO \cite{kopecky:91, kopecky:93, kopecky:98} formalisms assuming different $k_{0}$ enhancement factors.}
\end{figure}

The dominant decay following thermal neutron capture is by $E1$ primary $\gamma$-ray transitions.  The $E1$ photon strength is dominated by the low-energy tail of the giant dipole electric resonance (GDER). Theoretical models of the PSF describing the GDER are typically based on parametrizations of the corresponding giant resonance, observed in photonuclear reactions, whose transition probabilities are well described as a function of $\gamma$-ray energy \cite{krticka:08}.  Total photonuclear cross-section data derived from $^{186}$W photoabsorption measurements \cite{berman:69} can be used to test the validity for a variety of PSFs near the GDER.  These data ~\cite{berman:69} can be transformed to experimental PSF values $f^{(E1)}(E_{\gamma})$ using the empirical relationship of Ref.~\cite{kopecky:87}
\begin{equation}
f^{(E1)}(E_{\gamma}) = \frac{1}{3(\pi \hbar c)^{2}} \cdot \frac{\sigma_{\rm abs}}{E_{\gamma}},
\label{eq:insert2}
\end{equation}
where the constant $\frac{1}{3(\pi \hbar c)^{2}} = 8.68 \times 10^{-8}$~${\rm mb} \cdot {\rm MeV}^{-2}$, the photoabsorption cross section $\sigma_{\rm abs}$ is in units of [mb], and the $\gamma$-ray energy is in [MeV].  The results of this transformation for $^{186}$W are shown in Fig.~\ref{fig:2}.

\begin{table*}
\caption{\label{tab:3} GDER and GQER resonance parameters used in the {\small DICEBOX} simulations for the tungsten isotopes.  GDER parameters denoted by $G_{1}$ correspond to isovector vibrations along the symmetry axis, and parameters with $G_{2}$ correspond to isovector vibrations perpendicular to the symmetry axis.  The parametrizations for $^{183,185,187}$W are taken from nearest-neighboring even-odd isotope $^{189}$Os \cite{berman:79}, and the RIPL GDER parametrization corresponding to $^{186}$W was assumed for the even-even $^{184}$W.  The GQER parameters correspond to isovector-isoscalar vibrations based on a theoretical global parametrization.  See text for details.}
\begin{tabular}{|c|c|c|c|c|c|c|c|}\hline\hline
Isotope & Resonance & $E_{G_{1}}$ [MeV] & $\Gamma_{G_{1}}$ [MeV] & $\sigma_{G_{1}}$ [mb] & $E_{G_{2}}$ [MeV] & $\Gamma_{G_{2}}$ [MeV] & $\sigma_{G_{2}}$ [mb]\\
\hline

$^{183}$W & GDER & 12.68 & 2.71 & 268.0 & 14.68 & 3.62 & 395.0 \\
  & GQER & 11.10 & 3.91 & 4.55 & $-$ & $-$ & $-$ \\
$^{184}$W & GDER & 12.59 & 2.29 & 211.0 & 14.88 & 5.18 & 334.0 \\
  & GQER & 11.08 & 3.90 & 4.54 & $-$ & $-$ & $-$ \\
$^{185}$W & GDER & 12.68 & 2.71 & 268.0 & 14.68 & 3.62 & 395.0 \\
  & GQER & 11.06 & 3.89 & 4.53 & $-$ & $-$ & $-$ \\
$^{187}$W & GDER & 12.68 & 2.71 & 268.0 & 14.68 & 3.62 & 395.0 \\
  & GQER & 11.02 & 3.87 & 4.51 & $-$ & $-$ & $-$ \\
\hline\hline
\end{tabular}
\end{table*}

The Brink-Axel (BA) model \cite{brink:55,axel:62} and the enhanced generalized Lorentzian (EGLO) model \cite{kopecky:91, kopecky:93, kopecky:98} were used in these calculations to compare with experimental data.  The BA model is a form of the standard Lorentzian given by
\begin{equation}
f^{(E1)}_{\rm BA}(E_{\gamma}) = \frac{1}{3(\pi \hbar c)^{2}} \cdot \sum_{i=1}^{i=2} \frac{\sigma_{G_{i}} E_{\gamma} \Gamma_{G_{i}}^{2}}{(E_{\gamma}^{2}-E_{G_{i}}^{2})^{2} + E_{\gamma}^{2} \Gamma_{G_{i}}^{2}}.
\label{eq:BA}
\end{equation}
The resonance shape-driving parameters in Equation~\ref{eq:BA} are represented by the terms $E_{G_{i}}$ [MeV], the centroid of the GDER resonance, $\Gamma_{G_{i}}$ [MeV], the width of the resonance, and $\sigma_{G_{i}}$ [mb], the cross section of the resonance.  The adopted experimental parametrizations for the tungsten isotopes were taken from RIPL \cite{capote:09} and are listed in Table~\ref{tab:3}.  The corresponding BA PSF based on this parametrization is also shown in Fig.~\ref{fig:2} where it is compared to the experimental photoabsorption data.  Although these data are only available above $E_{\gamma} \gtrsim 9$~MeV, they demonstrate excellent agreement with the Brink hypothesis \cite{brink:55} in this region.

The EGLO model is derived from the idea of the generalized Lorentzian (GLO) model and was originally proposed by Kopecky and Uhl \cite{kopecky:90}, with the analytic form
\begin{eqnarray}
f^{(E1)}_{\rm GLO} (E_{\gamma}, \Theta) &=& \sum_{i=1}^{i=2} \frac{\sigma_{G_{i}} \Gamma_{G_{i}}}{3 (\pi \hbar c)^{2}} \left[ F_{K} \frac{4 \pi^{2} \Theta^{2} \Gamma_{G_{i}}}{E_{G_{i}}^{5}} \right. \nonumber \\
&& \left.+\frac{E_{\gamma} \Gamma_{G_{i}}(E_{\gamma}, \Theta)}{(E^{2}_{\gamma}-E^{2}_{G_{i}})^{2} + E^{2}_{\gamma} \Gamma^{2}_{G_{i}}(E_{\gamma}, \Theta)} \right].
\label{eq:7}
\end{eqnarray}
In this model a value of 0.7 has been used for the Fermi-liquid parameter $F_{K}$ \cite{kadmenski:83}.  This factor, together with the remaining terms of the first quotient in the parentheses of Equation~\ref{eq:7}, represent a correction to the Lorentzian function in describing the electric dipole operator in the limit of zero energy (as $E_{\gamma} \rightarrow 0$).  This form of the PSF is a violation of the Brink hypothesis since there is an additional dependence on the nuclear temperature $\Theta$, which may be written as a function of excitation energy
\begin{equation}
\Theta = \sqrt{(E_{\rm ex} - \Delta)/a},
\label{eq:8}
\end{equation}
where $E_{\rm ex}$ is the excitation energy of a final state, and $\Delta$ is the pairing energy.  The pairing correction has been determined according to the following convention: for even-even nuclei $\Delta = +0.5 \cdot |P_{d}|$ = 0.763 ($^{184}$W); for odd-$A$ nuclei $\Delta = 0$ ($^{183,185,187}$W); and for odd-odd nuclei $\Delta = -0.5 \cdot |P_{d}|$.  The deuteron-pairing energy, $P_{d}$ is tabulated in Ref.~\cite{vonegidy:05}.  Consequently, GDERs built on excited states may differ vastly in both shape and size to those built on the ground state since the width of the resonance is also a function of the nuclear temperature according to
\begin{equation}
\Gamma_{G_{i}} (E_{\gamma}, \Theta) = \frac{\Gamma_{G_{i}}}{E_{G_{i}}^{2}} (E_{\gamma}^{2} + 4 \pi^{2} \Theta^{2}).
\label{eq:9}
\end{equation}
In the EGLO version of this model, the term $\Gamma_{G_{i}}(E_{\gamma}, \Theta)$ has been modified by an enhancement factor given by an empirical generalization of the width \cite{kopecky:91, kopecky:93, kopecky:98}
\begin{equation}
\Gamma_{G_{i}}'(E_{\gamma},\Theta) = \left[ k_{0} + (1-k_{0}) \frac{(E_{\gamma} - E_{0})}{(E_{G_{i}} - E_{0})} \right] \Gamma_{G_{i}}(E_{\gamma},\Theta),
\label{eq:10}
\end{equation}
where $\Gamma_{G_{i}}'(E_{\gamma}, \Theta)$ is substituted for $\Gamma_{G_{i}}(E_{\gamma}, \Theta)$ in Equation~\ref{eq:7} to evaluate $f^{(E1)}_{\rm EGLO}(E_{\gamma}, \Theta)$.  A fixed value of $E_{0} = 4.5$~MeV has been adopted for the reference-energy \cite{kopecky:93, kopecky:98} and is found to have only a weak influence on the overall enhancement.  The parameter $k_{0}$ was then varied to optimize agreement with the absorption data of Ref.~\cite{berman:69}.  Figure~\ref{fig:2} shows that for $k_{0} = 3.5$ the EGLO PSF follows closely the experimental data for $E_{\gamma} \lesssim 17$~MeV.  Beyond this regime the PSF is heavily damped, however, these $\gamma$-ray energies are not of interest in thermal capture.  The GLO model is also plotted in Fig.~\ref{fig:2} along with an EGLO PSF using the empirically-determined value of $k_{0}$ from the mass-dependent model of Ref.~\cite{kopecky:98} where $k_{0} = 1+ \left[(0.09(A-148) \cdot {\rm exp}(-0.180(A-148))\right]$.  The plot illustrates very little difference in overall behavior between the GLO model and EGLO model with the mass-modeled-$k_{0}$ value.  Both PSFs fail to reproduce the experimental data at low energy and can only adequately describe the data in the double-humped resonance region.

For the magnetic-dipole transitions, $M1$, a PSF based on the single-particle (SP) model was adopted.  The value of $f^{(M1)}_{SP}$ was treated as an adjustable parameter in the {\small DICEBOX} calculations to obtain good agreement between statistical-model predictions and experimental-decay data in addition to the derived value of the total radiative capture width.  For the even-odd $^{183,185,187}$W compounds a value of $f^{(M1)}_{SP} = 1 \times 10^{-9}$~MeV$^{-3}$ was used, while a higher value of $f^{(M1)}_{SP} = 3 \times 10^{-9}$~MeV$^{-3}$ was found to reproduce the data better for the even-even $^{184}$W.  Other models, such as the scissors \cite{richter:90} and spin-flip \cite{b&m:75} models, were also be considered, however a lack of experimental evidence for a giant dipole magnetic resonance (GDMR) in the tungsten isotopes and the relative insignificance of these transitions in the calculations \cite{bolinger:70}, make the SP model a practical approach.

A giant quadrupole electric resonance (GQER) model has been used to describe the PSF for $E2$ multipoles.  This model is represented by a single-humped Lorentzian (cf. the standard Lorentzian in Equation~\ref{eq:BA}) to describe an isoscalar-isovector quadrupole-type vibration.  A global parametrization has been used to determine the set of resonance parameters, listed in Table~\ref{tab:3}.  The following convention was adopted in determining this parametrization: $E_{G} = 63 \cdot A^{-1/3}$~MeV \cite{speth:81}, $\Gamma_{G} = 6.11 - 0.012A$~MeV \cite{prestwitch:84}, and $\sigma_{G} = 1.5 \times 10^{-4} \cdot \frac{Z^{2}E_{G}^{2}A^{-1/3}}{\Gamma_{G}}$~mb \cite{prestwitch:84}.  Quadrupole strength contributes far less than the dipole strengths.  Transitions corresponding to higher multipoles, including $M2$, are not considered in modeling capture-state decay in this work.

\begin{figure}[b]
\includegraphics[angle=-90,width=\linewidth]{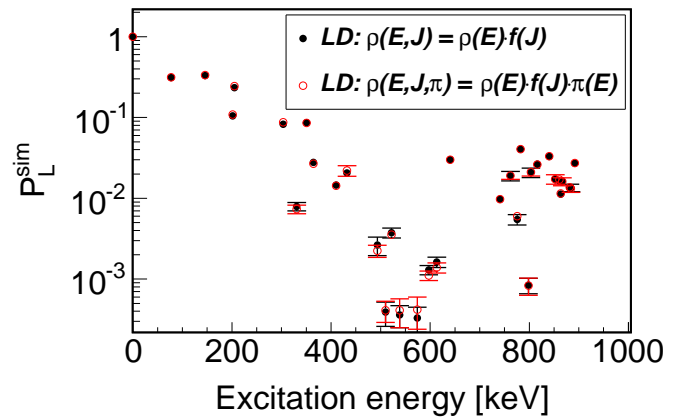}
\caption{\label{fig:LDPDEP} (Color online) Simulated populations to low-lying levels in $^{187}$W assuming a parity-independent (black) and parity-dependent (red) BSFG LD combined with the EGLO PSF.  The $\pi(E)$ dependence observed here is representative for all tungsten isotopes considered in this study.}
\end{figure}

\begin{table*}
\caption{\label{tab:183Wg} Experimental partial $\gamma$-ray cross sections, corresponding to both primary and secondary $\gamma$-ray transitions, measured in this work from thermal neutron capture on $^{182}$W.  Quantities in brackets represent tentative assignments.  Multipolarities, $XL$, in square brackets were assumed based on $\Delta J$ angular-momentum selection rules; other values were taken from ENSDF \cite{firestone:92}.}
\resizebox{\textwidth}{!}{
\begin{tabular}{c c c c c c c c c c c c}
\hline\hline

 $E_{\rm L}$ [keV] &  $J^{\pi}$ & $E_{\gamma}$ [keV] & $\sigma_{\gamma}^{\rm exp}$ [b] & $\alpha$ & $XL$ &$E_{\rm L}$ [keV] &  $J^{\pi}$ & $E_{\gamma}$ [keV] & $\sigma_{\gamma}^{\rm exp}$ [b] & $\alpha$  & $XL$ \\
\hline

\footnotetext[1]{Newly-placed transition based on statistical-model calculations.}
\footnotetext[2]{Multiplet resolved using ENSDF branching ratios \cite{ensdf}.}
\footnotetext[3]{Transition not observed in this work; intensity normalized to ENSDF-reported branching ratio \cite{ensdf}.}
\footnotetext[4]{Tentative $J^{\pi}$ assignment in ENSDF~\cite{ensdf}/RIPL~\cite{capote:09} confirmed by statistical-model calculations.}
\footnotetext[5]{The existence of this level is questionable, see text.}
\footnotetext[6]{Primary $\gamma$ ray observed by Bondarenko \textit{et al}. \cite{bondarenko:11}.}

0        & 1/2$^{-}$    &           &            &         &          &
         &             & 365.39(4)  & 0.0282(16) &  0.0474 & $E2$ \\

46.48    & 3/2$^{-}$    & 46.36(2)  & 1.078(92)  & 8.21    & $M1+E2$  &
453.07   & 7/2$^{-}$    & 40.976(1)\footnotemark[3]& 0.00187(21) & 11.4 &$M1$\\

99.08    & 5/2$^{-}$    & 52.52(2) & 0.305(24)   &  6.13     & $M1+E2$  &
         &             & 143.97(6) & 0.01041(89) & 1.72    & $M1+E2$ \\

         &             & 98.90(1)  & 0.342(12) & 4.05       & $E2$     & 
         &             & 161.17(5) & 0.0350(23) & 1.24      & $M1+E2$ \\

207.01   & 7/2$^{-}$    & 107.75(11) & 0.321(17) & 3.73     & $M1+E2$   &
         &             & 244.25(3)  & 0.0464(21) & 0.163    & $E2$ \\

         &             & 160.36(2) & 0.0995(61) & 0.661     & $E2$     &
         &             & 245.88(2) & 0.1069(54) & 0.385     & $M1+E2$ \\

208.81   & 3/2$^{-}$   & 109.55(1) & 0.1131(59) & 3.62       & $M1+E2$  &
         &             & 353.84(3) & 0.0494(23) & 0.139     & $M1+E2$ \\

         &             & 162.11(1) & 0.983(27)  & 1.15      & $M1+E2$  &
         &             & 406.23(7) & 0.0112(10) & 0.0355    & $[E2]$  \\

         &             & 208.64(2) & 0.1148(43)  & 0.527     & $M1+E2$  & 
475.21   & 11/2$^{-}$  & 166.39(15) & 0.00439(44) & 1.14     & $M1+E2$ \\

291.72   & 5/2$^{-}$   & 82.79(5)   & 0.0247(25)  & 8.24     & $M1+E2$  & 
         &            & 267.92(18) & 0.00435(93)  & 0.121    & $E2$ \\

         &            & 84.56(2)   & 0.0906(69) & 7.65      & $M1+E2$   &
485.10 & 13/2$^{+}$\footnotemark[4] & 175.89(1) & 0.0016(6) & 0.954 & $[M1]$\\

         &           & 192.49(3) & 0.0209(11) & 0.56        & $M1+E2$   &
(533)\footnotemark[5] & $(1/2, 3/2)$ & - & - & - & - \\ 

         &           & 245.24(1) & 0.0271(18) & 0.380       & $M1$      &
551.10   & (9/2$^{-}$) & 259.44(9) & 0.00788(96) & 0.134    & $E2$ \\

         &           & 291.57(1) & 0.2510(93) & 0.0926      & $E2$     & 
         &           & 344.02(13) & 0.0049(10) & 0.143      & $M1+E2$ \\

308.95   & 9/2$^{-}$ & (17.20(20))\footnotemark[1] & $1.08 \times 10^{-5}(4)$ & 16380 & $[E2]$  &
         &          & 452.37(9) & 0.00513(67) & 0.027       & $[E2]$  \\

       &        & 101.934(1)\footnotemark[2] & 0.00559(38) & 4.44 & $M1+E2$  &
595.3   & (9/2$^{-}$) & 142.11(4)  & 0.0174(11)   & 1.73       & $M1+E2$ \\

         &         & 209.69(2)  & 0.0756(29)   & 0.262      & $E2$     &
         &         & 286.39(1)\footnotemark[3] & 0.00052(26) & 0.249 & $[M1]$ \\

309.49   & 11/2$^{+}$ & 102.481(3)\footnotemark[2] & 0.0049(19) & 39.2 & $M2$ &
622.22   & 9/2$^{+}$ & 312.72(2)\footnotemark[2]  & 0.145(15)  & 0.199 & $M1$ \\

412.09   & 7/2$^{-}$ &  103.06(12) & 0.0147(92) & 4.35 & $M1$ & 
6190.88  & 1/2$^{+}$ & 5981.70(22)\footnotemark[6] & 0.0161(11)& 0 & $[E1]$ \\

         &          & 120.05(21) & 0.00202(71) & 2.12 & $[M1]$ & 
         &          & 6091.2(3)\footnotemark[6] & 0.0063(7) & 0 & $[M2]$ \\

         &          &  203.10(4) & 0.01711(92) & 0.298 & $E2$   &           
         &          & 6144.28(6) & 0.978(38) & 0 & $[E1]$ \\

         &          &  204.91(2) & 0.0484(20) & 0.630 & $M1+E2$ &
         &          & 6190.78(6) & 2.740 (38) & 0 & $[E1]$ \\

         &          & 313.02(5)\footnotemark[2] & 0.1833(91) & 0.194 & $M1+E2$&
         &          &  &  &  & \\
\hline\hline
\end{tabular}}
\end{table*}

\begin{figure*}
\includegraphics[angle=-90,width=\linewidth]{pd_w183_r.epsi}
\caption{\label{w183:PD} (Color online) Comparison of simulated population per neutron capture, given by {\small DICEBOX} ($P_{L}^{\rm sim}$), to experimental depopulation according to Equation~\ref{eq:14} ($P_{L}^{\rm exp}$), for low-lying levels below $E_{\rm crit}$ in $^{183}$W for various PSF/LD model combinations.  The spin distribution of low-lying levels is indicated in the upper panel of each plot, and the parity distribution for the same plot is shown in the lower panel.  For $E_{\rm crit} = 490$~keV in (a) and (b), good agreement between the statistical model and experimental data are attained, although the BA model does not reproduce the weakly-populated high-spin states as well as the EGLO model.  For $E_{\rm crit} = 625$~keV in (c) poorer agreement is observed, possibly due to missing levels above 490~keV.}
\end{figure*}

\section{\label{sec:level4}Results\protect\\}

\begin{figure}[t]
\includegraphics[angle=-90,width=\linewidth]{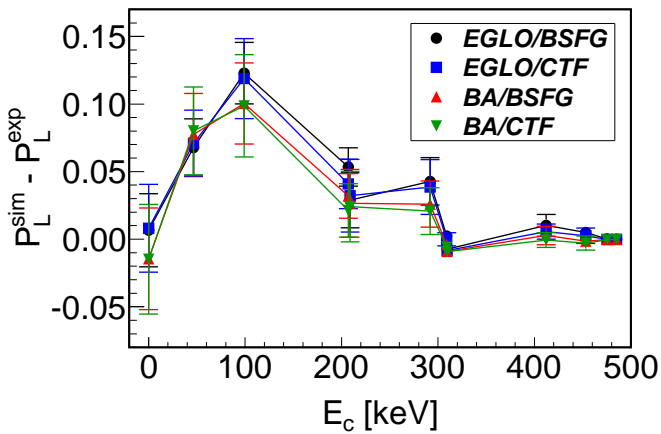}
\caption{\label{fig:deviation} (Color online) Comparison of the $^{183}$W simulated population and experimental depopulation, $P_{L}^{\rm sim} - P_{L}^{\rm exp}$, for different combinations of PSF/LD models as a function of $E_{\rm c}$.  The error bars show the systematic uncertainties in the {\small DICEBOX} calculations.  The point at 0~keV corresponds to the difference in GS feeding from the respective PSF/LD model calculations for $E_{\rm crit} = 490$~keV and the weighted average of the GS feeding of these PSF/LD combinations.}
\end{figure}

Thermal neutron-capture (n,$\gamma$) $\gamma$-ray cross sections depopulating levels in the $^{183,184,185,187}$W compounds, from irradiations of the isotopically-enriched $^{182,183,186}$W targets and a natural tungsten target for $^{184}$W(n,$\gamma$), are discussed below.  Only the primary $\gamma$ rays from the capture state or secondary $\gamma$ rays depopulating levels below $E_{\rm crit}$ are included in this paper.  The complete decay scheme determined in these measurements will be available in the EGAF database.

\begin{table}[b]
\caption{\label{tab:W182ng} Summary of $\sigma_{0}$ measurements for $^{182}$W(n,$\gamma$).}
\begin{tabular}{cc}
\hline\hline
$\sigma_{0}$ [b] & Reference \\
\hline
 {\bf 20.5(14)} & {\bf This work} \\
  19.2(19) & H.~Pomerance \cite{pomerance:52}  \\
  20.7(5)  & S.~J.~Friesenhahn \cite{friesenhahn:66} \\
  19.6(3)  & K.~Knopf \cite{knopf:87} \\
  20.0(6)  & V.~Bondarenko \cite{bondarenko:11} \\
  19.9(3)  & Atlas~\cite{mughabghab:06} \\
\hline\hline
\end{tabular}
\end{table}

All combinations of PSF and LD models described earlier, were used in the {\small DICEBOX} calculations and compared to experimental data  by plotting the simulated population against the experimental depopulation for each level below $E_{\rm crit}$ in population-depopulation plots.  For model combinations invoking the EGLO PSF we assumed a $k_{0} = 3.5$ enhancement factor.  Uncertainties in the population along the vertical axis correspond to Porter-Thomas fluctuations from independent nuclear realizations, while those along the horizontal axis are due to the experimental uncertainty in the measured cross sections depopulating the levels.  The vertical axis shows the calculated population per neutron capture to a given level, determined by {\small DICEBOX}, and the experimental depopulation of the corresponding level along the horizontal axis is normalized to the total radiative thermal-capture cross section according to
\begin{equation}
P_{L}^{\rm exp} = \sum\limits_{i=1}^{N} \frac{\sigma_{\gamma_{i}}(1+\alpha_{i})}{\sigma_{0}},
\label{eq:14}
\end{equation}
where $N$ denotes the number of $\gamma$ rays depopulating the level.

The population-depopulation plots compare the intensity balance through all states up to $E_{\rm crit}$. Scatter around the ${\rm population} = {\rm depopulation}$ line is a measure of the quality and completeness of the experimental data and provides a test of the ability of the statistical model to simulate the experimental decay scheme.  Model dependence in the population-depopulation plot is indicated by either smooth or spin dependent deviations, and isolated deviations for individual levels are indications of problems with the experimental $J^{\pi}$ assignments or other decay-scheme data.

In this work, we also investigated the parity dependence $\pi (E)$ on the overall LD assuming its separable form $\rho (E,J,\pi) = \rho(E) \cdot f(J) \cdot \pi(E)$.  The $\pi(E)$ dependence may be described by a Fermi-Dirac distribution parametrized according to Ref.~\cite{quraishi:03}.  In this framework, at large excitation energies $\pi(E) = 0.5$.  As $E \rightarrow 0$: $\pi(E) \rightarrow 1$ for even-even nuclei; $\pi(E) \rightarrow 0(1)$ for odd-$A$ nuclei for which the odd nucleon is in an odd-parity (even-parity) orbit; and, $\pi(E) \approx 0.5$ for odd-odd and odd-$A$ nuclei if the Fermi level is occupied by nearly degenerate positive- and negative-parity orbits.  Adopting an additional parity dependence in the LD models, $\rho (E,J,\pi) = \rho(E) \cdot f(J) \cdot \pi(E)$, the simulated populations for the odd-A isotopes $^{183,185,187}$W and even-even $^{184}$W were found to yield statistically consistent results with the parity-independent LD models, $\rho (E,J) = \rho(E) \cdot f(J)$; a representative comparison is illustrated in Fig.~\ref{fig:LDPDEP}.  A parity-independent approach was, therefore, considered adequate for modeling the LD in these analyses.

\subsection{\label{sec:level4.A}$^{182}$W(n,$\gamma$)$^{183}$W\protect\\}

A $^{182}$WO$_2$ target was irradiated for a 2.46-h period.  The current analysis and previous information in ENSDF \cite{firestone:92} implies that for $^{183}$W the level scheme is complete up to a level at 485.1~keV and we have set $E_{\rm crit} = 490.0$~keV, which includes an additional level over the value given in RIPL \cite{capote:09}.  A total of 12 levels in $^{183}$W are below $E_{\rm crit}$ with spins ranging from $1/2 \leq J \leq 13/2$, deexcited by 33 $\gamma$ rays and fed by four primary $\gamma$ rays, shown in Table~\ref{tab:183Wg}. Transition intensities have been corrected for absorption in the source, as discussed earlier. The multipolarities in Table~\ref{tab:183Wg} are taken from ENSDF \cite{firestone:92} where available, or assumed based on angular-momentum selection rules, and the conversion coefficients were recalculated with {\small BRICC} \cite{kibedi:08}.

Figure~\ref{w183:PD} shows the population-depopulation balance for $^{183}$W using the corresponding $\sigma_{\gamma}$ information from Table~\ref{tab:183Wg} calculated with various LD and PSF models.  These plots show little statistical-model dependence in the population of most excited states except for the high-spin $11/2^{+}$, $11/2^{-}$, and $13/2^{+}$ states at 309.5, 475.2, and 485.1~keV, respectively, that appear to be better reproduced using the EGLO PSF.  This is also shown in Fig.~\ref{fig:deviation} where the difference in the {\small DICEBOX}-modeled population ($P_{L}^{\rm sim}$) for a variety of PSF/LD combinations and the experimental depopulation ($P_{L}^{\rm exp}$) is model independent and insensitive to cut-off energies, $E_{\rm c}$, above 300 keV.  Figure~\ref{fig:deviation} shows excellent consistency between the models at each value of $E_{\rm c}$.

\begin{figure}[t]
\includegraphics[angle=-90,width=\linewidth]{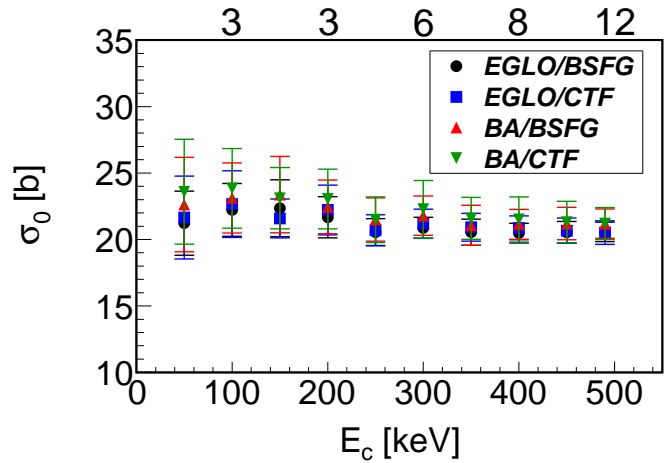}
\caption{\label{fig:w183sigma0} Variation of the total radiative thermal neutron-capture cross section ($\sigma_{0}$) with increasing cut-off energy for the reaction $^{182}$W(n,$\gamma$)$^{183}$W assuming different combinations of PSF/LD models.  The numbers along the top axis indicate the cumulative number of known experimental levels at the corresponding value of $E_{\rm c}$.  The error bars only show systematic uncertainties from the {\small DICEBOX} calculations.}
\end{figure}

The total-capture cross section, $\sigma_{0}$, determined for the different PSF/LD combinations, is also independent of $E_{\rm crit}$ for various model combinations as seen in Fig.~\ref{fig:w183sigma0}.  For $E_{\rm crit} = 100$~keV, with only three low-lying levels, $\sigma_{0}$ remains nearly constant although the systematic uncertainty is larger.  This rapid convergence is due to the dominant ground-state feeding from experimental transitions deexciting low-lying levels that dominates the calculation.  We adopt the value $\sigma_{0} = 20.5(14)$~b corresponding to the EGLO/CTF combination.  Of the $\sim 7$~\% uncertainty on our value, the systematic uncertainty from the simulated cross section is 4.3~\% and $\gamma$-ray self attenuation accounts for 3.2~\%. The statistical and normalization errors are far less significant with each only contributing $\lesssim 2$~\%.  The result for the total radiative thermal-capture cross section for $^{182}$W(n,$\gamma$)$^{183}$W is consistent with the recommended value of 19.9(3)~b \cite{mughabghab:06} and previous experimental investigations \cite{pomerance:52,friesenhahn:66,knopf:87,bondarenko:11} listed in Table~\ref{tab:W182ng}.

The choice of PSF and LD combination has a pronounced effect on the calculated capture-state total radiative width. The EGLO/CTF result, $\Gamma_{0} = 0.040(3)$~eV, agrees best with the recommended value of $\langle \Gamma_{0} \rangle = 0.051(4)$~eV.  For the EGLO/BSFG and BA/CTF combinations somewhat poorer agreement is obtained with $\Gamma_{0}$ values of 0.071(3) and 0.076(6) eV respectively.  The BA/BSFG combination gives much poorer agreement with $\Gamma_{0}=0.138(7)$ eV.  Fortunately, the choice of PSF/LD model has only a small effect on the derived cross section.

\begin{table*}
\caption{\label{tab:184Wg} Experimental partial $\gamma$-ray cross sections, corresponding to both primary and secondary $\gamma$-ray transitions, measured in this work from thermal neutron capture on $^{183}$W.  Quantities in brackets represent tentative assignments.  Multipolarities, $XL$, in square brackets were assumed based on $\Delta J$ angular-momentum selection rules; other values were taken from ENSDF \cite{baglin:10}.}
\resizebox{\textwidth}{!}{
\begin{tabular}{cccccccccccc}
\hline\hline
 $E_{\rm L}$ [keV] &  $J^{\pi}$ & $E_{\gamma}$ [keV] & $\sigma_{\gamma}^{\rm exp}$ [b] & $\alpha$ & $XL$ & $E_{\rm L}$ [keV] &  $J^{\pi}$ & $E_{\gamma}$ [keV] & $\sigma_{\gamma}^{\rm exp}$ [b] & $\alpha$ & $XL$ \\
\hline

\footnotetext[1]{Tentative $J^{\pi}$ assignment in ENSDF~\cite{ensdf}/RIPL~\cite{capote:09} confirmed by statistical-model calculations.}
\footnotetext[2]{Transition not observed in this work; intensity normalized to ENSDF-reported branching ratio~\cite{ensdf}.}
\footnotetext[3]{Doublet resolved using ENSDF-reported branching ratios \cite{ensdf}.}
\footnotetext[4]{The existence of this level is questionable, see text.}
\footnotetext[5]{Newly-placed transition based on statistical-model calculations.}
\footnotetext[6]{Newly-identified $\gamma$ rays based on experimental observation.}
\footnotetext[7]{Primary $\gamma$ ray observed by Bushnell \emph{et al}. \cite{bushnell:75}.}

0       & $0^{+}$  &               &          &          &          &
1252.20 & $8^{+}$  & 504.03(20)    & $<0.00016$ & 0.0206 & $E2$\\

111.22  & $2^{+}$  & 111.19(2) & 1.597(44)   & 2.59   & $E2$   &
(1282.71)\footnotemark[4] & $(1^{-},2^{-})$   & - & - & - & - \\

364.07  & $4^{+}$ & 252.86(1) & 0.714(31) & 0.145 & $E2$ &
1285.00 & $5^{-}$ & 63.689(1)\footnotemark[2] & 0.00141(15) & 25.75 & $E2$\\

748.32  & $6^{+}$ & 384.08(8) & 0.0242(16) & 0.0419 & $E2$ &
        &        & 151.13(2)\footnotemark[2] & 0.000147(20) & 0.1286 & $E1$\\

903.31  & $2^{+}$ & 539.40(23) & 0.0106(24) & 0.0175 & $E2$ &
        &         & (279.0)\footnotemark[2] & $<2.58 \times 10^{-6}$ & 1.111 & $[M2]$\\

      &         & 792.09(2) & 1.157(50) & 0.00733 & $M1+E2$ & 
       &        & 381.82(14)\footnotemark[2] & 0.000178(24) & 0.1579 & $[E3]$\\

      &         & 903.31(3) & 1.185(52) & 0.00554 & $E2$ &
      &         & 536.79(22) & 0.0094(28) & 0.00612 & $E1+M2$ \\

1002.49 & $0^{+}$ & 891.30(2) & 0.596(26) & 0.0057 & $[E2]$ &
      &         & 921.01(9) & 0.0258(26) & 0.0030 & $E1+M2$ \\

1005.97 & $3^{+}$ & 641.79(8) & 0.0850(48) & 0.01183 & $M1+E2$ &
       &       & 1173.77(3)\footnotemark[2] & 0.00384(42) & 0.000698 & $[E3]$\\

       &         & 894.78(2) & 0.686(30) & 0.00569 & $M1+E2$ &
1294.94 & $5^{+}$ & (9.94)\footnotemark[5] & 0.00250(56) & 8.829 & $[E1]$ \\

1121.44 & $2^{+}$ & 757.37(3) & 0.220(10) & 0.00804 & $E2$ &
       &           & 930.76(23) & 0.0094(21) & 0.0116 & $[M1]$\\

       &         & 1010.26(3) & 0.346(16) & 0.0139 & $M1+E2$ &
1322.15 & $0^{+}$\footnotemark[1] & 418.88(2) & 0.0062(11) & 0.0333 & $[E2]$\\

       &         & 1121.32(4) & 0.1360(85) & 0.00359 & $E2$ &
       &         & 1211.0(10)\footnotemark[3] & 0.0059(29) & 0.00310 & $[E2]$\\

1130.05 & $2^{-}$\footnotemark[1] & 124.04(2) & 0.0579(34) & 0.215 & $[E1]$ & 
1345.37 & $4^{-}$\footnotemark[1] & 211.61(16) & 0.0062(11) & 0.0547 & $E1$\\

       &         & 226.75(1) & 0.694(30) & 0.059 & $E1+M2$ &
       &         & (215.21(10)) & 0.0015(70) & 0.242 & $E2$\\

       &         & 1018.68(9) & 0.0437(30) & 0.0017 & $[E1]$ &
       &         & 339.48(2) & 0.0340(16) & 0.0170 & $[E1]$ \\

1133.85 & $4^{+}$ & 127.61(6)\footnotemark[2] & 0.000173(71) & 1.57 & $M1+E2$ &
       &         & 981.1(5)\footnotemark[2] & 0.0051(20) & 0.00185 & $[E1]$ \\

       &          & 230.45(6)\footnotemark[2] & 0.00152(24) & 0.1932 & $E2$ &
1360.38 & $4^{+}$\footnotemark[1] & (65.36(19)\footnotemark[6]) & 0.0135(26) & 2.792 & $[E1]$ \\

       &        & 385.5\footnotemark[2] & $<0.000574$ & 0.0414 & $[E2]$ &
       &          & 238.52(25) & 0.0036(12) & 0.174 & $[E2]$ \\

        &        & 769.78(2)\footnotemark[3] & 0.0692(56) & 0.0080 & $M1+E2$ & 
       &          & 996.04(6)\footnotemark[3] & 0.0180(78) & 0.00977 & $M1$ \\

       &         & 1022.58(9) & 0.0512(32) & 0.0043 & $E2$ &
       &        & 1249.8(10)\footnotemark[2] & 0.00158(68) & 0.00292 & $[E2]$\\

1221.31 & $3^{-}$ & 87.34(6) & 0.0138(17) & 0.533 & $E1$ &
7411.11 & $0^{-},1^{-}$ & 6089.1(3)\footnotemark[7] & 0.0061(5) & 0 & $[E1]$ \\

       &         & 91.17(12) & 0.0052(12) & 0.603 & $M1+E2$ &
        &        & 6281.5(4)\footnotemark[7] & 0.0101(9) & 0 & $[M1]/[E2]$ \\

       &         & 215.34(3)\footnotemark[3] & 0.0959(51) & 0.0521 & $E1$ &
       &         & 6289.51(13) & 0.214(13) & 0 & $[E1]$ \\

       &         & 318.03(2) & 0.1961(87) & 0.0202 & $E1+M2$ &
       &         & 6408.6(12) & 0.395(21) & 0 & $[E1]$ \\

       &         & 857.73(21) & 0.0077(24) & 0.0024 & $E1$ &
       &         & 6507.63(16) & 0.0852(60) & 0 & $[E1]$ \\

       &         & 1109.72(20) & 0.0283(34) & 0.0016 & $E1+M2$ & 
       &         & 7299.69(16) & 0.1353(85) & 0 & $[E1]$ \\

      &        & 1221.29(4)\footnotemark[2] & 0.000706(67) & 0.0064& $[E3]$ & 
       &         & 7410.99(14) & 0.535(25) & 0 & $[E1]$ \\

\hline\hline
\end{tabular}}
\end{table*}

\begin{figure*}
\includegraphics[angle=-90,width=\linewidth]{pd_w184_r.epsi}
\caption{\label{w184:PD} (Color online) The simulated population per neutron capture, given by {\small DICEBOX} ($P_{L}^{\rm sim}$), versus the experimental depopulation according to Equation~\ref{eq:14} ($P_{L}^{\rm exp}$), for low-lying levels beneath $E_{\rm crit} = 1370.0$~keV in $^{184}$W, assuming EGLO/BSFG and EGLO/CTF model combinations.  The spin distribution of low-lying levels is indicated in the upper panel of each plot, and the parity distribution for the same plot is shown in the lower panel.  The plots give the best agreement between the statistical model and experimental data in (a) and (b) where the $1^{-}$ capture-state spin is dominant.  As the $0^{-}$ capture state becomes increasingly dominant in (c) agreement between simulation and experiment becomes notably worse.  The levels most-adversely affected by the increase in the $0^{-}$ contribution are circled.}
\end{figure*}

The $11/2^{+}$ ($T_{1/2}=5.2$~s) isomer at 309.49~keV \cite{firestone:92} decays by a highly-converted 102.48-keV \cite{firestone:92} $M2$ transition that was not resolved from the 101.93-keV transition deexciting the 308.95-keV level and the 101.80-keV transition deexciting the 302.35-keV level in $^{187}$W which also contributes to the observed intensity due to a 0.5(1)~\% $^{186}$W impurity (Table~\ref{tab:isocomp}) in the measured sample.  The total intensity of the triplet is $\sim 15(2)$~\% of the 209.69-keV $\gamma$-ray intensity deexciting the 308.95-keV level, which is significantly larger than 7.4(4)~\% observed from the same level in $^{183}$Ta $\beta^{-}$ decay~\cite{firestone:92}.  Assuming the excess intensity, after the additional correction for the $^{186}$W impurity (see Section~\ref{sec:level4.D}), comes from the isomer transition, we get $\sigma_{\gamma}(102.48)=0.0049(19)$~b.  Accounting for internal conversion this gives an experimental depopulation of 0.197(76)~b which is consistent with the observed total $\gamma$-ray intensity feeding the metastable isomer, $\sigma_{11/2^{+}}(^{183}{\rm W}^{m}) = 0.177(18)$~ b, from the 485.72- and 622.22-keV levels which are deexcited by transitions at 175.89 and 312.72~keV, respectively.  The combined intensity of these transitions yields $\sum\sigma_{\gamma}^{\rm exp}(11/2^{+}) = 0.177(18)$~b and the DICEBOX-modeled population of the 309.49-keV isomer is $P(11/2^{+})=0.00154(97)$.  The experimental depopulation of the 309.49-keV level is consistent with the simulated population from our {\small DICEBOX} calculations to within 3~$\sigma$ as indicated in the log-log space of Fig.~\ref{w183:PD}.  The current measurement supports the proposed $J^{\pi} = 13/2^{+}$ assignment for the 485.72-keV level that was previously reported in reaction experiments \cite{saitoh:00}.  Our simulations also support the inclusion of a new, highly-converted, 17.2-keV $E2$ transition deexciting the 308.95-keV level with a total intensity of $\sim 180$~mb feeding the 291.72-keV level that improves the agreement between population and depopulation for both levels. The 17.2-keV transition is below the detection threshold of our HPGe detector.

The next level above $E_{\rm crit}$ at 533~keV is reported in ENSDF \cite{firestone:92} with $J^{\pi} = (1/2,3/2)$.  The 533-keV level was only reported as populated by primary $\gamma$-rays in a resonance (n,$\gamma$) experiment \cite{casten:73} and not seen in our work or later (n,$\gamma$) or reaction experiments.  The existence of this level is considered doubtful; certainly the proposed $J^{\pi}$ assignment is highly questionable since these states are expected to be strongly populated in $s$-wave capture on $^{182}$W (see Fig.~\ref{w183:PD}).  Raising the cut-off energy to 625-keV and including the next three levels at 551.1, 595.3, and 622.22~keV leads to poorer agreement in the population-depopulation balance for several levels as shown in Fig.~\ref{w183:PD}(c).  We observe the transitions from these three levels, but since the statistical model gives better agreement for $E_{\rm crit} = 490$ keV, it is likely that the decay-scheme information is incomplete between the 490 and 622.22~keV.

\subsection{\label{sec:level4.B}$^{183}$W(n,$\gamma$)$^{184}$W\protect\\}

\begin{table}[b]
\caption{\label{tab:183Wr} Summary of $^{183}$W(n,$\gamma$) $\sigma_{0}$ measurements.}
\begin{tabular}{cc}
\hline\hline
$\sigma_{0}$ [b] & Reference \\ \hline
  {\bf{9.37(38)}} & {\bf This work} \\
  10.9(11) & H.~Pomerance \cite{pomerance:52}  \\
  10.0(3)  & S.~J.~Friesenhahn \cite{friesenhahn:66} \\
  10.5(2)  & K.~Knopf \cite{knopf:87} \\
  10.4(2)  & Atlas ~\cite{mughabghab:06} \\
\hline\hline
\end{tabular}
\end{table}

\begin{figure}[t]
\includegraphics[angle=-90,width=\linewidth]{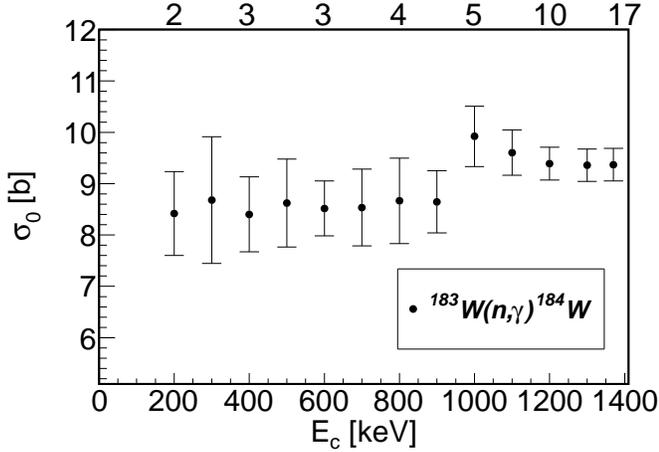}
\caption{\label{fig:sigma0Ec_w184} Variation of the total radiative thermal neutron-capture cross section ($\sigma_{0}$) with increasing cut-off energy for $^{183}$W(n,$\gamma$) using the EGLO/CTF combination and assuming the capture-state composition $J^{\pi} = 0^{-}(7.4~\%)+1^{-}(92.6~\%)$.  The numbers along the top axis indicate the cumulative number of known experimental levels at the corresponding value of $E_{\rm c}$.  The error bar only shows the systematic contribution from the {\small DICEBOX} calculations.}
\end{figure}

A $^{183}$WO$_2$ target was irradiated for 2.24 h.  Comparison of the DICEBOX-population calculations with the experimental depopulation data for $^{184}$W sets $E_{\rm crit} = 1370.0$~keV.  This value is higher than in RIPL where $E_{\rm crit} = 1252.2$~keV and includes 12 levels.  There are 18 levels below our cut-off energy including one tentative level assignment.  The $^{184}$W decay scheme consists of seven primary $\gamma$-rays and 47 secondary $^{184}$W $\gamma$-rays that are listed in Table~\ref{tab:184Wg}.  The experimental multipolarities and mixing ratios are taken from ENSDF \cite{baglin:10} where available or assumed based on selection rules.  The ground state of the $^{183}$W target nucleus is $J^{\pi} = 1/2^{-}$, allowing $s$-wave neutron capture to populate resonances with $J^{\pi} = 0^{-},1^{-}$.  The \emph{Atlas of Neutron Resonances} \cite{mughabghab:06} indicates that $1^{-}$ capture-states account for 78.3~\% of the observed total-capture cross section, 7.4~\% is from $0^{-}$ capture states, and the remaining 14.3~\% of the cross section is attributed to a negative-parity {\sl bound} resonance at $E_{0} = -26.58$~eV (with respect to the separation energy) with unknown spin.

\begin{figure}[t]
\includegraphics[angle=-90,width=\linewidth]{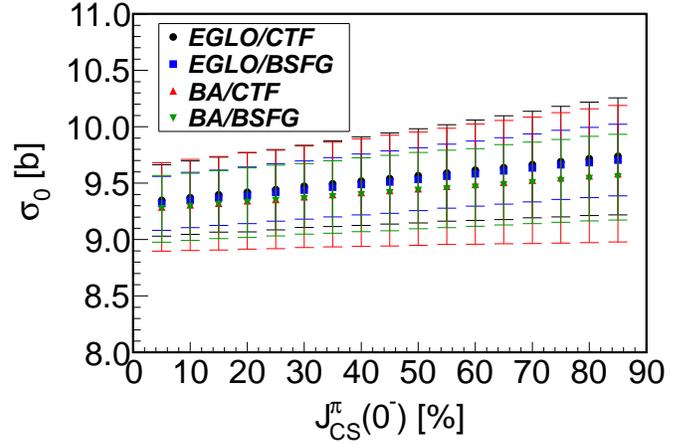}
\caption{\label{fig:s0_varyCS} (Color online) The effect on $\sigma_{0}$ of varying the relative capture-state spin composition $0^{-} + 1^{-}$ in $^{184}$W, assuming different combinations of PSF/LD.  The uncertainty corresponds to the modeled cross section only.}
\end{figure}

\begin{figure}[b]
\includegraphics[angle=-90,width=\linewidth]{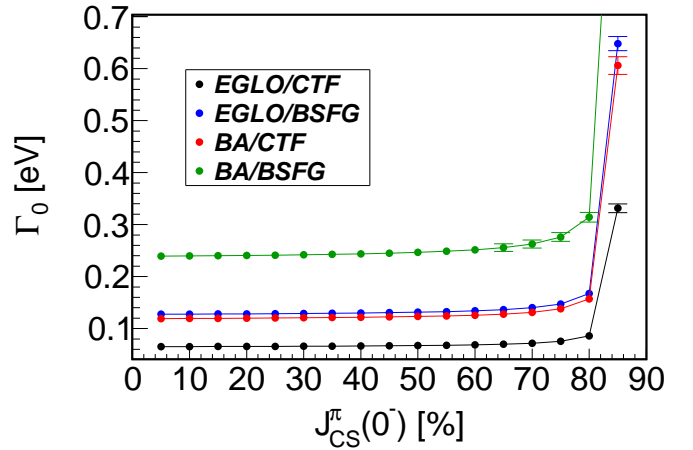}
\caption{\label{fig:Gamma0_CS} (Color online) The effect on $\Gamma_{0}$ of varying the relative capture-state $0^{-}$+$1^{-}$ spin composition in $^{184}$W assuming different combinations of PSF/LD models.}
\end{figure}

The population-depopulation plots in Figs.~\ref{w184:PD}(a) and (b) show that $\sigma_0$ is insensitive to both the $0^{-}/1^{-}$ composition of the capture state and the choice of PSF and LD combinations.  Figure~\ref{fig:sigma0Ec_w184} shows the dependence of the derived cross section on $E_{\rm c}$.  For $E_{\rm c} \leq 900$~keV there are only four levels and $\sigma_0 = 8.65(64)$~b.  Adding the level at 903.31~keV, which feeds the ground state with $\sigma_{\gamma} = 1.185(52)$~b, increases the derived cross section significantly, demonstrating the necessity to include as many experimentally known low-lying levels as possible in the simulation.  For $E_{\rm crit}$ = 1370.0~keV, with a total of 17 levels (not including the tentative 1282.7-keV level, see later), we get $\sigma_{0} = 9.37(38)$~b, which is comparable at 2 $\sigma$ with the recommended value of 10.4(2)~b \cite{mughabghab:06} and previous measurements shown in Table~\ref{tab:183Wr}.  We also find that the total thermal-capture cross section is statistically insensitive to the $J^{\pi}$ composition of the capture state as illustrated in Fig.~\ref{fig:s0_varyCS}.  The overall uncertainty on our adopted value for $\sigma_{0}$ of 4.0~\% is dominated by the 3.4~\% systematic uncertainty in the simulation and the 1.7~\% statistical uncertainty.  Uncertainties due to $\gamma$-ray self attenuation and normalization are much lower, each contributing $ < 1.0$~\%.

\begin{figure}[t]
\includegraphics[angle=-90,width=\linewidth]{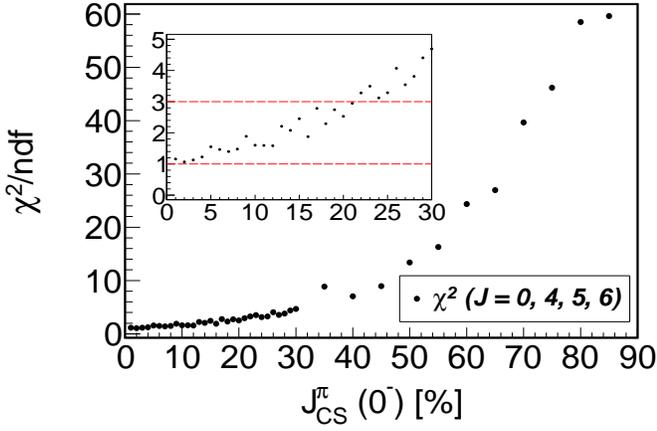}
\caption{\label{fig:chi2_JPi} (Color online) Reduced $\chi^{2}$ calculated using Equation~\ref{eq:chi2_highJ} as a summation over weakly-populated low-lying levels: 748.32~keV, $6^{+}$; 1133.85~keV, $4^{+}$; 1285.00~keV, $5^{-}$; 1294.94~keV, $5^{+}$; 1322.15~keV, $0^{+}$; 1345.37~keV, $4^{-}$; and, 1360.38~keV, $4^{+}$, assuming different $J^{\pi}$ capture-state compositions.  The inset figure is expanded around $0~\%\leq J^{\pi}(0^{-}) \leq 30~\%$.  The statistical Porter-Thomas fluctuations are apparent in the plot.  Dashed-red lines are drawn at values of $\chi^{2}/{\rm ndf}$ of 1.0 and 3.0, respectively, in the inset.}
\end{figure}

The capture-state width, $\Gamma_{0}$, is strongly dependent on the choice of PSF/LD combination, but is only weakly influenced by the capture-state spin composition, as shown in Fig.~\ref{fig:Gamma0_CS}: $\Gamma_{0}$ is nearly constant up to $\sim 65$-\% $0^{-}$ contribution, and only gradually increases up to $\sim 80~\%$.  The EGLO/CTF model combination, with a 78.3-\% $1^-$ capture-state composition (Fig.~\ref{w184:PD}(b)), gives $\Gamma_0$=0.066(2)~eV, in agreement with the adopted value of 0.073(6)~eV \cite{mughabghab:06}.  For the model combinations: EGLO/BSFG, $\Gamma_{0} = 0.129(3)$; BA/CTF, $\Gamma_{0} = 0.121(3)$; and BA/BSFG, $\Gamma_{0}=0.242(6)$; all are substantially higher than the adopted value.  The effect of the capture-state composition is most sensitive to the modeled population of the $0^{+}$ and $J \geq 4$ low-lying levels. For $0^{-}$ capture-state compositions of 7.4~\% (Fig.~\ref{w184:PD}(a)) and 21.7~\% (Fig.~\ref{w184:PD}(b)), the EGLO results give excellent agreement with experiment.  If the $0^{-}$ capture-state composition increases to 85~\% (Fig.~\ref{w184:PD}(c)), the predicted population of $0^{+}$ and high-spin states is much poorer.  The 85-\% $0^{-}$ composition also gives $\Gamma_{0}$ values of 0.348(8) for the EGLO/BSFG model combination and 0.178(5)~eV for the EGLO/CTF combination that are considerably higher than the adopted value.  To determine the most likely $J^{\pi}$ capture-state composition we varied this parameter and calculated the corresponding reduced $\chi^{2}$, using the population-depopulation data for the weakly populated states (circled in Fig.~\ref{w184:PD}), as 
\begin{equation}
\chi^{2}/{\rm ndf} = \sum \frac{(P_{L}^{\rm exp} - P_{L}^{\rm sim})^{2}}{(dP_{L}^{\rm sim})^{2}},
\label{eq:chi2_highJ}
\end{equation}
where $P_{L}^{\rm exp}$ is the expectation value.  Figure~\ref{fig:chi2_JPi} shows that $\chi^{2}$ approaches 1.0 for capture-state compositions with $J^{\pi}(0^{-}) < 10$~\%.  Indeed, the simulated populations to these levels is more than $3~\sigma$ away from the expectation value assuming $J^{\pi}(0^{-}) \approx 22$~\%.  This result implies a likely capture-state composition $J^{\pi}(0^{-}) \lesssim 7$~\%, and hence, $J^{\pi} = 1^{-}$ is the most probable assignment for the bound resonance at $-26.58$~eV \cite{mughabghab:06}.  Thus, an overall fractional distribution of $J^{\pi} = 0^{-}(7.4~\%) + 1^{-}(92.6~\%)$ is consistent with the capture-state composition of Ref.~\cite{mughabghab:06}.

\begin{figure}[t]
\includegraphics[angle=-90,width=\linewidth]{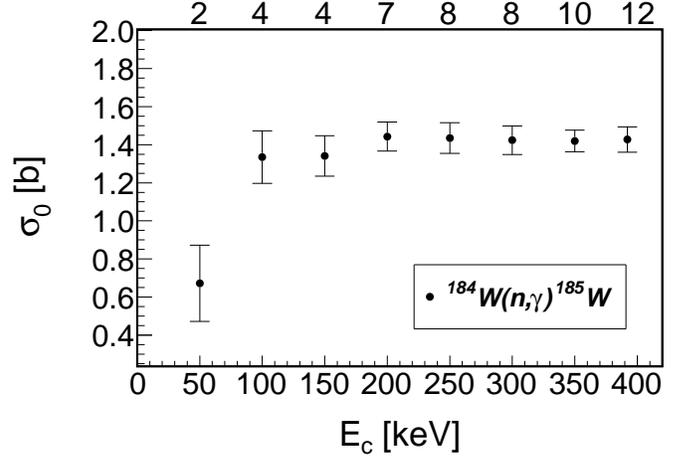}
\caption{\label{fig:sigma0Ec_w185} Variation of the total radiative thermal neutron-capture cross section ($\sigma_{0}$) with increasing cut-off energy for $^{184}$W(n,$\gamma$) using the EGLO/BSFG combination.  The numbers along the top axis indicate the cumulative number of known experimental levels at the corresponding value of $E_{\rm c}$.  The error bar only shows the systematic contribution from the {\small DICEBOX} calculations.}
\end{figure}

Our analysis confirms the decay scheme for $^{184}$W reported in ENSDF~\cite{baglin:10} except for the 161.3-keV $\gamma$ ray depopulating the 1282.71-keV $(1,2)^{-}$ level, which we did not observe.  This level assignment was tentative and the 161.3-keV $\gamma$-ray was placed twice in the level scheme (also depopulating the $6^{-}$ level at 1446.27~keV).  Since this level is expected to be strongly populated, we conclude that it most likely does not exist (or has a considerably different $J^{\pi}$) and have removed it from our analysis.  We have also assigned a new $\gamma$ ray at 65.36(19)~keV, depopulating the 1360.38-keV level. Another 9.94-keV $\gamma$ ray depopulating the 1294.94-keV level is proposed based on the population-depopulation balance.  The 504.03-keV $\gamma$ ray deexciting the 1252.20-keV $8^{+}$ level was not firmly identified although we can set an experimental limit of $\sigma_{\gamma} < 0.16$~mb which is consistent with statistical-model predictions of 0.1(1)~mb.

\begin{figure*}
\includegraphics[angle=-90,width=\linewidth]{pd_w185_r.epsi}
\caption{\label{w185:PD} (Color online) The simulated population per neutron capture, given by {\small DICEBOX} ($P_{L}^{\rm sim}$), versus the experimental depopulation according to Equation~\ref{eq:14} ($P_{L}^{\exp}$), for low-lying levels below $E_{\rm crit} = 392.0$~keV in $^{185}$W assuming the EGLO/BSFG model combinations.  The spin distribution of low-lying levels is indicated in the upper panel of each plot, and the parity distribution for the same plot is shown in the lower panel.  Poor agreement between the simulations and experiment for the 187.88- and 390.92-keV levels in (a), where both levels are over-populated cf. experiment, is improved by the addition of low-energy $\gamma$-ray transitions depopulating these levels.}
\end{figure*}

Some $\gamma$ rays from levels below $E_{\rm crit}$ were not observed in our data and their relative cross sections were taken from ENSDF \cite{baglin:10}, normalized to the cross sections of (observed) stronger transitions from those levels, as indicated in Table~\ref{tab:184Wg}. An unresolved doublet centered at 769~keV $\gamma$-ray deexcites the 1133.85- and 1775.34-keV levels and was resolved using the ENSDF-adopted branching intensities from both levels.  Doublets centered around 215 and 996~keV, depopulating levels at 1221.31 and 1360.38~keV, respectively, were also resolved in a similar manner, as indicated in Table~\ref{tab:184Wg}.  The 1285.00-keV level is an 8.33-$\mu$s isomer with $J^{\pi}=5^{-}$, and is populated with a cross section $\sigma_{5^{-}} = 24.7(55)$~mb from beneath $E_{\rm crit}$; transitions from above $E_{\rm crit}$ known to feed the isomer were not observed in this work.

\subsection{\label{sec:level4.C}$^{184}$W(n,$\gamma$)$^{185}$W\protect\\}

\begin{table*}
\caption{\label{tab:185Wg} Experimental partial $\gamma$-ray cross sections, corresponding to both primary and secondary $\gamma$-ray transitions, in $^{185}$W.  The energies are from Ref.~\cite{bondarenko:05}.  Intensities from Ref.~\cite{bondarenko:05} were normalized to cross sections using data measured on a natural tungsten sample in this work.  Quantities in brackets represent tentative assignments.  Multipolarities, $XL$, in square brackets were assumed based on $\Delta J$ angular-momentum selection rules; other values were taken from ENSDF \cite{wu:05}.}
\resizebox{\textwidth}{!}{
\begin{tabular}{cccccccccccc}
\hline\hline
 $E_{\rm L}$ [keV] &  $J^{\pi}$ & $E_{\gamma}$ [keV] & $\sigma_{\gamma}^{\rm exp}$ [b] & $\alpha$ & $XL$ & $E_{\rm L}$ [keV] &  $J^{\pi}$ & $E_{\gamma}$ [keV] & $\sigma_{\gamma}^{\rm exp}$ [b] & $\alpha$ & $XL$ \\
\hline

\footnotetext[1]{Transition not reported in Ref.~\cite{bondarenko:05}; intensity estimated based on statistical-model calculations.}
\footnotetext[2]{Transition not reported in Ref.~\cite{bondarenko:05}; intensity normalized to ENSDF-reported branching ratio \cite{ensdf}.}
\footnotetext[3]{Newly-placed transition based on statistical-model calculations.}
\footnotetext[4]{Transition not reported in Ref.~\cite{bondarenko:05}; intensity calculated assuming feeding to the 1.67-min isomer at 197.38~keV $\sigma_{11/2^{+}}=0.0062(16)$~b$-$see text.}
\footnotetext[5]{Tentative $J^{\pi}$ assignment in ENSDF~\cite{ensdf}/RIPL~\cite{capote:09} confirmed by statistical-model calculations.}

0       & $3/2^{-}$ &          &         &           &        &
        &          & 150.3(2)\footnotemark[2] & 0.00014(4) & 0.834 & $[E2]$\\

23.55   & $1/2^{-}$ & 23.54(4)\footnotemark[1] & 0.00266(13) & 90 & $[M1+E2]$ &
        &          & 177.36(6)& 0.0286(25) & 0.89 & $M1+E2$\\

65.85   & $5/2^{-}$  & 42.29(5)\footnotemark[2] & $2.31 \times 10^{-5}(71)$ & 189 & $E2$ & 
        &          & 243.38(15) & 0.00460(44) & 0.162 & $[E2]$\\

        &           & 65.86(3)\footnotemark[1] & 0.0242(20) & 13 & $M1+E2$ & 
301.13 & $9/2^{-}$\footnotemark[5] & 127.4(5) & 0.00128(42) & 2.37 & $[M1+E2]$\\

93.30  & $3/2^{-}$ & 93.30(5)\footnotemark[1] & 0.0417(66) & 5.79 & $[M1+E2]$ &
       &          & 235.17(11) & 0.00371(44) & 0.423 & $[M1+E2]$\\

173.70 & $7/2^{-}$ & 107.85(2)& 0.0082(13) & 3.24 & $M1+E2$ &
332.11 & $7/2^{-}$\footnotemark[5]& 144.16(7)& 0.00371(56)& 1.66 & $[M1+E2]$ \\

       &          & 173.68(2) & 0.0676(48) & 0.499 & $E2$ &
       &          & 158.29(14) & 0.00140(63) & 1.278 & $[M1+E2]$ \\

187.88 & $5/2^{-}$ & (14.16(6))\footnotemark[3] & $2.82 \times 10^{-4}(15)$ & 254.8 & $[M1]$ &
       &          & 238.74(7)  & 0.00193(73) & 0.172 & $[E2]$ \\

       &          & 94.59(4) & 0.00315(55) & 5.4 & $[M1+E2]$ &
       &          & 266.24(7)  & 0.00972(81  & 0.301 & $[M1+E2]$ \\

       &          & 122.05(7) & 0.00402(57) & 2.3 & $[M1+E2]$ &
381.70 & $(13/2^{+})$ & - & - & - & - \\

       &          & 164.33(2) & 0.0189(14) & 0.606 & $[E2]$ & 
390.92 & $9/2^{-}$\footnotemark[5] & (58.37(20))\footnotemark[3] & 0.00084(13) & 3.889 & $[M1]$\\

        &           & 187.88(2) & 0.0258(20) & 0.59 & $[M1+E2]$ & 
       &          & 147.08(6) & 0.00104(16) & 1.57 & $[M1+E2]$\\

197.38 & $11/2^{+}$ & 23.54(5)\footnotemark[4] & $5.1 \times 10^{-8}(21)$ & 8339 & $[M2]$ &
5753.74 &  $1/2^{+}$ & 5658.6(11)  & 0.00176(71) & 0 & $[E1]$\\

       &           & 131.55(2)\footnotemark[4] & $2.82 \times 10^{-4}(71)$ & 19.39 & $E3$ & 
       &            & 5729.2(13)  & 0.00130(61) & 0 & $[E1]$\\

243.62 & $7/2^{-}$ & 69.7(3) & 0.0022(3) & 3.3 & $M1+E2$ &
       &            & 5753.65(5)  & 0.0546(38)  & 0 & $[E1]$\\

\hline\hline

\end{tabular}}
\end{table*}

A $^{\rm nat}$WO$_{2}$ target was irradiated for 11.52 h.  Comparison of the DICEBOX-population calculations with the experimental-depopulation data for the $^{185}$W compound sets $E_{\rm crit} = 392.0$~keV.  This value is higher than in RIPL where $E_{\rm crit} = 243.4$~keV which includes eight levels. Table~\ref{tab:185Wg} lists 11 levels beneath the cut-off energy, deexcited by 25 secondary $\gamma$ rays, and populated by three primary $\gamma$ rays.  These data were measured with a natural tungsten sample and supplemented with data from Bondarenko \textit{et al}.~\cite{bondarenko:05} that was renormalized to our cross sections.  Ten levels below $E_{\rm crit}$ have negative parity with spins ranging from $1/2^{-}$ to $9/2^{-}$, and there are two positive-parity levels at 197.43 (11/2$^{+}$, $T_{1/2}=1.67$~min) \cite{wu:05} and 381.70~keV (13/2$^{+}$) \cite{bondarenko:05} that are high-spin with no $\gamma$ rays observed deexciting them.  We have used the total cross section populating the 197.43-keV level from higher-lying levels in $^{185}$W from Ref.~\cite{bondarenko:05}, $\sigma_{11/2^{+}} = 6.2(16)$~mb, to determine the $\gamma$-ray cross sections deexciting this isomer.  This cross section is substantially larger that than the recommended value, $\sigma_0 = 2(1)$~mb \cite{mughabghab:06}.  The positive-parity levels below $E_{\rm crit}$ play only a small role in our simulations and do not limit the choice of $E_{\rm crit}$.  The mixing ratios and multipolarities in Table~\ref{tab:185Wg} were taken from ENSDF \cite{wu:05} where available or assumed based on selection rules associated with the $\Delta J$ transitions.

We determined the thermal-capture cross section, $\sigma_{0} = 1.43(10)$~b, for $^{184}$W(n,$\gamma$).  The result is largely insensitive with respect to PSF/LD combinations and comparable to the adopted value $\sigma_{0} = 1.7(1)$~b \cite{mughabghab:06}.  Table~\ref{tab:184Wr} shows the comparison of our value with other reported measurements.  For the EGLO/BSFG model combination, shown in Fig.~\ref{fig:sigma0Ec_w185}, $\sigma_{0}$ is statistically independent of $E_{\rm crit}$.  The uncertainty in $\sigma_{0}$ is 7~\%.  Several low-energy $\gamma$ rays contribute significantly to $\sigma_{0}$ but were not observed by experiment and were, instead, estimated from statistical-model calculations.  The systematic uncertainty in the ground-state feeding from the simulation is 4.7~\%.  A statistical uncertainty of 3.2~\%  and an uncertainty of 2.4~\% in the normalization also contribute.  The data from Ref.~\cite{bondarenko:05} were measured with a very thin target so no correction due to $\gamma$-ray self attenuation was required.

The total radiative width of the capture state in $^{185}$W varies widely depending on the choice of PSF/LD models.  The EGLO/BSFG combination generates a total width $\Gamma_{0} = 0.052(3)$~eV that is in excellent agreement with the adopted value, $\langle \Gamma_{0} \rangle = 0.052(4)$~eV \cite{mughabghab:06}.  Other combinations show poorer agreement: $\Gamma_{0} = 0.034(3)$~eV for EGLO/CTF; $\Gamma_{0} = 0.069(6)$~eV for BA/CTF; and, $\Gamma_{0} = 0.108(7)$~eV for the BA/BSFG combination.

\begin{table}[b]
\caption{\label{tab:184Wr} Summary of $^{184}$W(n,$\gamma$) $\sigma_{0}$ measurements.}
\begin{tabular}{cc}\hline\hline
$\sigma_{0}$ [b] & Reference \\ \hline
 {\bf 1.43(10)} & {\bf This work} \\
  2.12(42) & L.~Seren \cite{seren:47} \\
   1.97(30) & H.~Pomerance \cite{pomerance:52}  \\
   2.28(23) & W.~S.~Lyon \cite{lyon:60} \\
   1.70(10) & S.~J.~Friesenhahn \cite{friesenhahn:66} \\
   1.70(10) & K.~Knopf \cite{knopf:87} \\
   1.76(9)  & V.~Bondarenko \cite{bondarenko:05} \\
   2.40(10) & V.~A.~Anufriev \cite{anufriev:83} \\
   1.70(10) & Atlas~\cite{mughabghab:06} \\
\hline\hline
\end{tabular}
\end{table}

Here we report more precise energies for the 301.13 and 332.11-keV levels than are in ENSDF~\cite{wu:05}.  No $\gamma$ rays were previously reported deexciting these levels. Our {\small DICEBOX} calculations support the results of Bondarenko \emph{et al}. \cite{bondarenko:05} where six new $\gamma$ rays were identified depopulating these levels.  Two new, low-energy $\gamma$ rays are proposed deexciting levels at 187.88 ($E_{\gamma} \approx 14$~keV) and 390.92~keV ($E_{\gamma} \approx 58$~keV) based on the population-depopulation intensity balance.  The $\sim 58$-keV $\gamma$-ray transition is highly 
\begin{longtable*}{cccccccccccc}
\caption{\label{tab:187Wg} Experimental partial $\gamma$-ray cross sections, corresponding to both primary and secondary $\gamma$-ray transitions, measured in this work from thermal neutron capture on $^{186}$W.  Quantities in brackets represent tentative assignments.  Multipolarities, $XL$, in square brackets were assumed based on $\Delta J$ angular-momentum selection rules; other values were taken from ENSDF \cite{basunia:09}.}\\
\hline\hline

 $E_{\rm L}$ [keV] &  $J^{\pi}$ & $E_{\gamma}$ [keV] & $\sigma_{\gamma}^{\rm exp}$ [b] & $\alpha$ & $XL$ & $E_{\rm L}$ [keV] &  $J^{\pi}$ & $E_{\gamma}$ [keV] & $\sigma_{\gamma}^{\rm exp}$ [b] & $\alpha$ & $XL$ \\
\hline\endfirsthead

\caption{\emph{continued}}\\
\hline\hline
 $E_{\rm L}$ [keV] &  $J^{\pi}$ & $E_{\gamma}$ [keV] & $\sigma_{\gamma}^{\rm exp}$ [b] & $\alpha$ & $XL$ & $E_{\rm L}$ [keV] &  $J^{\pi}$ & $E_{\gamma}$ [keV] & $\sigma_{\gamma}^{\rm exp}$ [b] & $\alpha$ & $XL$ \\
\hline\endhead

\multicolumn{12}{r}{\emph{continued on next page}}
\endfoot
\hline\hline \endlastfoot

\footnotetext[1]{Tentative $J^{\pi}$ assignment in ENSDF~\cite{ensdf}/RIPL~\cite{capote:09} confirmed by statistical-model calculations.}

\footnotetext[2]{Multiplet transition resolved using ENSDF-reported branching ratios \cite{ensdf}.}

\footnotetext[3]{Newly-placed transition based on statistical-model calculations.}
\footnotetext[4]{New $J^{\pi}$ assignment based on statistical-model calculations.}
\footnotetext[5]{Transition inferred by coincidence data~\cite{bondarenko:08}; cross section deduced from observed intensity feeding the 401.06-keV level and statistical-model predictions.}

\footnotetext[6]{Weak evidence for transition in this work; intensity normalized to ENSDF-reported branching ratio \cite{ensdf}.}

\footnotetext[7]{Newly-identified level.}
\footnotetext[8]{Newly-identified $\gamma$ ray based on experimental observation.}

\footnotetext[9]{Multiplet transition resolved using experimental data and statistical-model calculations.}

\footnotetext[10]{Primary $\gamma$ ray observed by Bondarenko {\it et al}. \cite{bondarenko:08}.}

0 & $3/2^{-}$     &                 &            &            & &
(493.4)\footnotemark[7] & $(9/2^{-})$\footnotemark[4] & (143.2(1))\footnotemark[9] & 0.0222(58)& 1.7 & $[M1]$ \\

77.29 & $5/2^{-}$ & 77.30(5) & 0.823(14) & 10.17 & $M1+E2$ &
510.00 & $11/2^{-}$\footnotemark[1] & (145.8(1))\footnotemark[9] & 0.0052(21) & 1.62 & $[M1]$ \\

145.85 & $1/2^{-}$ & 145.84(5) & 4.727(46) & 1.65 & $M1$ & 
522.15 & $9/2^{-}$\footnotemark[1] & 171.70(6) & 0.0526(32) & 0.71 & $[M1+E2]$\\

201.45 & $7/2^{-}$ & 124.18(5) & 0.282(16) & 2.01 & $M1+E2$ & 
538.45 & $11/2^{-}$\footnotemark[1] & 337.18(19) & 0.0096(18) & 0.0604 &$[E2]$\\

       &           & 201.51(5) & 1.515(76) & 0.303 & $[E2]$ &
574.05 & $11/2^{-}$\footnotemark[1] & 209.59(33) & 0.0042(16) & 0.59 & $[M1]$\\

204.90 & $3/2^{-}$ & 59.30(5) & 1.048(43) & 3.73 & $M1$ &
597.24 & $11/2^{+}$\footnotemark[4] & - & - & - & \\

       &           & 127.55(5) & 0.646(37) & 1.87 & $M1+E2$ &
613.38 & $9/2^{-}$\footnotemark[1] & (16.20(13))\footnotemark[3] & (0.00293(43)) & 10.08 & $[E1]$\\

       &           & 204.87(5) & 0.666(33) & 0.631 & $[M1]$ & 
       &             & 282.86(19) & 0.0054(18) & 0.259 & $[M1]$\\

303.35 & $5/2^{-}$ & 98.51(8) & 0.0261(27) & 4.97 & $[M1]$ &
       &             & 310.52(12) & 0.0119(17) & 0.0771 & $[E2]$\\

       &           & 101.80(5) & 0.234(16) & 4.61 & $M1$ &
       &             & 410.8(5)\footnotemark[2] & 0.00031(5) & 0.0944 & $[M1]$\\

       &           & 157.47(5) & 0.1474(81) & 0.713 & $[E2]$ &
640.49 & $5/2^{-}$ & 276.19(6) & 0.0635(48) & 0.109 & $[E2]$\\

       &           & 226.02(5) & 0.379(19) & 0.243 & $M1+E2$ &
       &          & 289.98(6) & 0.300(15) & 0.17 & $M1+E2$\\

       &           & 303.31(6) & 0.248(13) & 0.213 & $[M1]$ & 
       &           & 438.91(10) & 0.0174(24) & 0.0794 & $[M1]$\\

330.78 & $9/2^{-}$\footnotemark[1] & 129.1(2)\footnotemark[2] & 0.0051(38) & 2.34 & $[M1]$ &
       &           & 563.33(13) & 0.0321(49) & 0.0415 & $[M1]$\\

       &           & 253.51(5)\footnotemark[2] & 0.1268(92) & 0.143 & $[E2]$ &
       &           & 640.55(10) & 0.085(10) & 0.0298 & $[M1]$ \\

350.43 & $7/2^{-}$ & (19.60(5))\footnotemark[3] & 0.00051(18) & 97.66 & $[M1]$&
710.78 & $13/2^{-}$\footnotemark[1] & (380.0(2))\footnotemark[9] & (0.00030(30)) & 0.0431 & $[E2]$\\

       &           & 148.89(5) & 0.204(11) & 1.55 & $[M1]$ &
727.86 & $11/2^{-}$\footnotemark[1] & 205.7(1)\footnotemark[2] & 0.0016(4) & 0.6166 & $[M1]$\\

       &           & 273.12(5) & 1.337(14) & 0.283 & $[M1]$ &
       &           & 377.0(2)\footnotemark[2] & 0.0014(2) & 0.0444 & $[E2]$\\

       &           & 350.34(9) & 0.0219(25) & 0.0542 & $[E2]$ &
741.08 & $7/2^{+}$\footnotemark[1] & 218.81(7) & 0.0220(27) & 0.0503 & $[E1]$\\

364.22 & $9/2^{-}$ & (13.80(4))\footnotemark[3] & 0.00293(22) & 275.2 & $[M1]$&
       &           & 330.97(6) & 0.0775(45) & 0.1682 & $[M1]$\\

       &           & 162.59(12) & 0.0100(15) & 1.2 & $[M1]$ &
       &           & 376.80(5)\footnotemark[2] & 0.184(21) & 0.0134 & $[E1]$\\

       &           & 286.79(7) & 0.0314(26) & 0.0981 & $[E2]$ &
       &           & 390.56(10) & 0.0661(42) & 0.0123 & $[E1]$\\

410.06 & $9/2^{+}$\footnotemark[4] & 45.8(3)\footnotemark[5] & 0.301(12) & 0.5941 & $[E1]$ &
       &           & 539.58(14)\footnotemark[2] & 0.0092(21) & 0.00605 &$[E1]$\\

432.28 & $7/2^{-}$ & 128.93(6)\footnotemark[2] & 0.1064(82) & 2.34 & $[M1]$ &
       &           & 663.91(8) & 0.0764(62) & 0.00394 & $[E1]$\\

       &           & 227.37(10) & 0.0506(37) & 0.203 & $[E2]$ &
762.15 & $1/2^{-}$\footnotemark[1] & 557.24(5)\footnotemark[2] & 0.572(33)\footnotemark[2] & 0.0427 &$(M1+E2)$\\

       &           & 230.56(14) & 0.0148(33) & 0.453 & $[M1]$ &
       &             & 616.33(5) & 0.304(16) & 0.0329 & $[M1]$\\

       &           & 354.92(7) & 0.1814(95) & 0.14 & $[M1]$ &
       &             & 762.0(5)\footnotemark[2] & 0.0286(60) & 0.0191 & $[M1]$\\

       &           & 432.4(5)\footnotemark[6] & 0.0098(28) & 0.0305 & $[E2]$ &
775.60 & $7/2^{-}$\footnotemark[1] & (135.1(5))\footnotemark[8] & 0.0478(71) & 2.01 & $[M1]$\\


       &           & 253.50(16)\footnotemark[2] & 0.0205(47) & 0.349 & $[M1]$ &
       &           & 659.18(9)\footnotemark[2] & 0.0738(91) & 0.0109 & $[E2]$\\

       &           & 411.28(9) & 0.0164(23) & 0.0944 & $[M1]$ &
       &           & 783.74(13) & 0.0836(82) & 0.0179 & $[M1]$\\

 782.29 & $1/2^{-}$ & 577.36(5)\footnotemark[2] & 0.921(46) & 0.031 &$(M1+E2)$ &
       &           & 860.77(12) & 0.1058(82) & 0.0141 & $[M1]$\\

       &           & 636.64(35)\footnotemark[2] & 0.0396(42) & 0.0303 & $[M1]$&
863.29 & $5/2^{-}$\footnotemark[1] & 513.0(5)\footnotemark[2] & 0.024(10) & 0.0531 & $[M1]$\\

       &           & 704.9(4)\footnotemark[2] & 0.0138(20) & 0.0094 & $[E2]$ &
       &           & 532.41(7)\footnotemark[2] & 0.039(12) & 0.0180 & $[E2]$ \\

       &           & 782.25(5) & 0.606(31) & 0.0179 & $[M1]$ &
       &           & 559.79(9) & 0.0283(32) & 0.0423 & $[M1]$\\

797.03 & $11/2^{-}$\footnotemark[1] & 364.7(1)\footnotemark[9] & 0.0013(5) & 0.0482 & $[E2]$ &
       &              & 658.0(3)\footnotemark[2] & 0.0168(84) & 0.0279 & $[M1]$\\

       &              & 466.3(1)\footnotemark[9] & 0.0017(5) & 0.0680 & $[M1]$ &
       &              & 661.9(3)\footnotemark[2] & 0.038(15) & 0.0275 & $[M1]$\\

798.22 & $(9/2^{+})$ & - & - & - & - &
       &             & 717.36(14) & 0.0300(64) & 0.00905 & $[E2]$\\

803.37 & $3/2^{-}$\footnotemark[1] & 500.02(6)\footnotemark[2] & 0.115(17) & 0.0565 & $(M1)$ &
       &             & 785.73(11) & 0.0850(81) & 0.0177 & $[M1]$\\

       &             & 598.55(15)\footnotemark[2] & 0.0608(88) & 0.0355 & $[M1]$ &
       &             & 862.96(10)\footnotemark[2] & 0.099(11) & 0.014 & $[M1]$\\

       &             & 657.50(7)\footnotemark[2] & 0.320(32) & 0.0279 & $[M1]$ &
866.68 & $3/2^{-}$\footnotemark[1] & 563.51(6)\footnotemark[2] & 0.023(13) & 0.0157 & $[E2]$\\

       &             & 726.03(5) & 0.1118(74) & 0.0216 & $[M1]$ &
       &             & 661.65(7)\footnotemark[2] & 0.068(24) & 0.0275 & $[M1]$\\

       &             & 803.25(8)\footnotemark[2] & 0.1043(70) & 0.0168 & $[M1]$ &
       &             & 789.38(10) & 0.234(52) & 0.00735 & $[E2]$\\

809.79 & $(13/2^{-})$ & - & - & - & - &
       &             & 866.37(13) & 0.278(16) & 0.0139 & $[M1]$\\

811.7  & $(15/2^{+})$ & - & - & - & - &
881.77 & $5/2^{+}$\footnotemark[1] & 140.47(13) & 0.0260(55) & 1.82 & $[M1]$\\

815.51 & $13/2^{+}$\footnotemark[4] & - & - & - & - &
       &             & 449.58(11)\footnotemark[2] & 0.0086(43) & 0.00899 & $[E1]$\\

816.26 & $3/2^{-}$ & 176.6(6)\footnotemark[6] & 0.0087(46) & 0.9436 & $[M1]$ &
       &           & 531.29(5)\footnotemark[2] & 0.201(26) & 0.00624 & $[E1]$\\

       &           & 383.87(8) & 0.0217(22) & 0.0422 &$[E2]$&
       &           & 676.79(8) & 0.0475(50) & 0.00379 & $[E1]$\\

       &           & 465.54(8)\footnotemark[2] & 0.0464(34) & 0.0252 & $[E2]$ &
       &           & 679.97(14)\footnotemark[2] & 0.0105(45) & 0.00375 &$[E1]$\\

       &           & 512.52(14)\footnotemark[2] & 0.065(13) & 0.0531 &$[M1]$&
       &             & 803.7(4)\footnotemark[2] & 0.0171(15) & 0.0027 & $[E1]$\\

       &           & 611.34(5)\footnotemark[2] & 0.167(21) & 0.0336 & $(M1)$ &
       &           & 881.58(6) & 0.214(12) & 0.00226 & $[E1]$\\

       &           & 670.37(5) & 0.227(12) & 0.0265 & $[M1]$ &
884.13 & $(5/2^{+})$\footnotemark[4] & 143.15(6)\footnotemark[2] & 0.0414(36) & 1.71 & $[M1]$\\

       &           & 738.84(6) & 0.185(10) & 0.0208 & $[M1]$ &
       &           & 243.63(37)\footnotemark[6] & 0.00296(15) & 0.0381 & $[E1]$\\

       &           & 816.20(20) & 0.436(67) & 0.0161 & $[M1]$ &
       &           & 451.29(19)\footnotemark[2] & 0.0065(21) & 0.0089 & $[E1]$\\

840.21 & $1/2^{-}$\footnotemark[1] & 537.21(23) & 0.0114(41) & 0.0176 &$[E2]$&
       &             & 474.02(6) & 0.296(15) & 0.0240 & $[E2]$\\

       &             & 635.37(8)\footnotemark[2] & 0.1059(86) & 0.0304 & $[M1]$ &
       &             & 533.63(6) & 0.0934(64) & 0.00619 & $[E1]$\\

       &             & 694.33(5) & 0.235(13) & 0.0243 & $[M1]$ &
891.93 & $3/2^{-}$\footnotemark[1] & 460.1(8)\footnotemark[6] & 0.0069(25) & 0.0259 & $[E2]$\\

       &             & 762.82(7) & 0.172(14) & 0.00792 & $[E2]$ &
       &             & 541.46(7) & 0.0848(64) & 0.0173 & $[E2]$\\

       &             & 840.17(5) & 0.662(34) & 0.015 & $[M1]$ &
       &             & 588.55(6) & 0.0971(62) & 0.0371 & $[M1]$\\

852.41 & $3/2^{-}$ & 502.0(6)\footnotemark[2] & 0.0137(60) & 0.0209 & $[E2]$ &
       &          & 690.15(16)\footnotemark[2] & 0.0082(41) & 0.00985 & $[E2]$\\

       &           & 549.0(5)\footnotemark[2] & 0.0195(80) & 0.0443 & $[M1]$&
       &           & 745.88(5) & 0.236(13) & 0.0203 & $[M1]$\\

       &           & 647.41(8) & 0.1065(73) & 0.029 & $[M1]$ &
       &           & 814.03(19)\footnotemark[2] & 0.122(10) & 0.0162 & $[M1]$\\

       &           & 650.88(14) & 0.0212(41) & 0.0113 & $[E2]$ &
       &           & 891.89(5)\footnotemark[2] & 0.408(22) & 0.0129 & $[M1]$\\

       &           & 706.59(6)\footnotemark[2] & 0.195(16) & 0.0232 & $[M1]$ &
5466.62 & $1/2^{+}$  & 4574.67(7) & 0.397(21) & 0 & $[E1]$\\

       &           & 774.92(6)\footnotemark[2] & 0.128(13) & 0.0184 & $[M1]$ &
       &           & 4585.7(6)\footnotemark[10] & 0.0052(20) & 0 & $[E2]$ \\

       &           & 852.18(6) & 0.160(11) & 0.0144 & $[M1]$ &
       &           & 4602.6(15)\footnotemark[10] & 0.024(12) & 0 & $[E1]$ \\

860.76 & $3/2^{-}$\footnotemark[1] & 428.48(8) & 0.0701(48) & 0.0313 & $[E2]$ &
        &          & 4606.6(11)\footnotemark[10] & 0.0159(60) & 0 & $[E1]$ \\

       &           & 655.87(7) & 0.227(14) & 0.0281 & $[M1]$ &
        &          & 4615.3(7)\footnotemark[10] & 0.0052(12) & 0 & $[E1]$ \\


        &          & 4626.40(7) & 0.627(33) & 0 & $[E1]$ &
        &          & 5163.5(4)\footnotemark[10] & 0.0135(20) & 0  & $[M2]$\\

        &          & 4650.27(8) & 0.207(12) & 0 & $[E1]$ &
        &          & 5388.85(26)\footnotemark[10] & 0.0143(12) & 0 & $[M2]$\\

        &          & 4662.94(27) & 0.0197(30) & 0 & $[E1]$ &
        &          & 5261.67(9) & 2.297(32) & 0 & $[E1]$\\

        &          & 4684.31(7) & 0.765(40) & 0 & $[E1]$ &
        &          & 5466.47(12) & 0.0675(50) & 0 & $[E1]$\\

       &           & 4704.8(4)\footnotemark[10] & 0.0091(12) & 0 & $[E1]$ &
       &           & 5320.65(8) & 1.625(83) & 0 & $[E1]$\\

       &           & 4826.0(10)\footnotemark[10] & 0.0048(12) & 0 & $[M2]$ &
       &           &            &           &   & \\

\end{longtable*}
\noindent
converted and obscured by a strong tungsten X~ray at 57.98~keV, making a $\gamma$ ray of this energy difficult to observe.  Both new transitions were assumed to have $M1$ multipolarity.  The improvement by including these transitions is shown in Fig.~\ref{w185:PD}.  The $^{185}$W $\gamma$ rays deexciting the first three excited states at 23.55, 65.85, and 93.30~keV were not observed in either this work or that of Bondarenko \emph{et al}. \cite{bondarenko:05}.  The transition cross sections depopulating these levels were determined from the simulated cross section populating those levels,  using the EGLO/BSFG model combination and the branching ratios from ENSDF \cite{wu:05}, as shown in Fig.~\ref{w185:PD}.  Our {\small DICEBOX}-simulated population per neutron capture to each of these levels is: 23.55~keV, 0.178(28); 65.85~keV, 0.254(33); and, 93.30~keV, 0.201(32).  These values can be compared to those of Bondarenko \emph{et al}. \cite{bondarenko:05}: 23.55~keV, 0.168(16); 65.85~keV, 0.126(14); and, 93.30~keV, 0.201(17).  The difference between simulation and Ref.~\cite{bondarenko:05} for the 65.85-keV level implies there is a substantial contribution from the quasi continuum that is not observed experimentally.  Four levels were previously reported with tentative $J^{\pi}$ assignments \cite{wu:05}.  For three of these levels, our simulations are consistent with the assignments of $9/2^{-}$, $7/2^{-}$, and $9/2^{-}$ to the 301.13-, 332.11-, and 390.4-keV levels, respectively.  The agreement between modeled population and experimental depopulation by assuming these $J^{\pi}$-level assignments is illustrated in the population-depopulation plot of Fig.~\ref{w185:PD}(b).  Those assignments are also consistent with the distorted-wave Born approximation (DWBA) calculations described in Ref.~\cite{bondarenko:05}.

\begin{figure}[b]
\includegraphics[angle=-90,width=\linewidth]{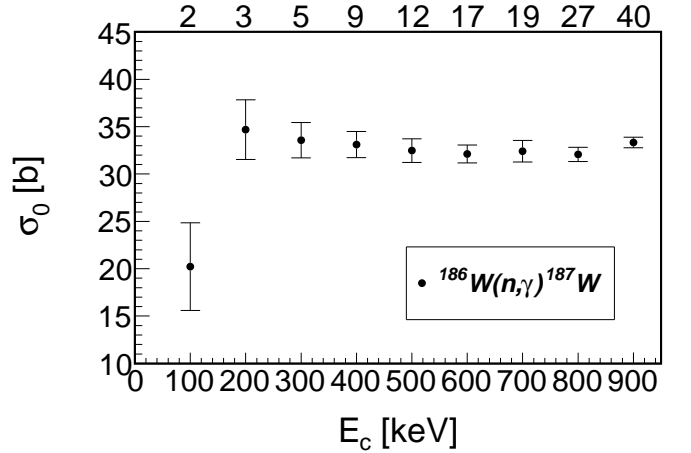}
\caption{\label{fig:sigma0Ec_w187} Variation of the total radiative thermal neutron-capture cross section ($\sigma_{0}$) with increasing cut-off energy for $^{186}$W(n,$\gamma$) using the EGLO/BSFG combination.  The numbers along the top axis indicate the cumulative number of known experimental levels at the corresponding value of $E_{\rm c}$.  The error bar only shows the systematic contribution from the {\small DICEBOX} calculations.}
\end{figure}

\begin{figure*}
\includegraphics[angle=-90,width=\linewidth]{pd_w187_r_A.epsi}
\caption{\label{w187:PD} (Color online) The simulated population per neutron capture, given by {\small DICEBOX} ($P_{L}^{\rm sim}$), versus the experimental depopulation according to Equation~\ref{eq:14} ($P_{L}^{\rm exp}$), for low-lying levels below $E_{\rm crit} = 900.0$~keV in $^{187}$W assuming the EGLO/BSFG model combination.  The spin distribution of low-lying levels is indicated in the upper panel of each plot, and the parity distribution for the same plot is shown in the lower panel.  Excellent agreement is seen over five orders of magnitude except for poor agreement between simulation and experiment (a) for the level at 364.22~keV where {\small DICEBOX} predicts a higher population than is observed experimentally. Improvement between model and experiment is attained (b) by the addition of a low-energy $\sim 14$-keV $\gamma$ transition in the decay of the 364.22-keV level.}
\end{figure*}

\begin{figure*}
\includegraphics[angle=-90,width=\linewidth]{pd_w187_r_B.epsi}
\caption{\label{w187:PD_2} (Color online) The simulated population per neutron capture, given by {\small DICEBOX} ($P_{L}^{\rm sim}$), versus the experimental depopulation according to Equation~\ref{eq:14} ($P_{L}^{\rm exp}$), for low-lying levels below $E_{\rm crit} = 900.0$~keV in $^{187}$W assuming the EGLO/BSFG model combination.  The spin distribution of low-lying levels is indicated in the upper panel of each plot, and the parity distribution for the same plot is shown in the lower panel.  In (a) poor agreement between {\small DICEBOX} calculations and experiment is attained assuming $J^{\pi}$ assignments of $7/2^{+}$ and $11/2^{+}$ for the 884.13- and 410.06-keV levels, respectively.  Excellent agreement is seen in (b) assuming $J^{\pi}$ assignments of $5/2^{+}$ and $9/2^{+}$ for the 884.13- and 410.06-keV levels, respectively.}
\end{figure*}

\subsection{\label{sec:level4.D}$^{186}$W(n,$\gamma$)$^{187}$W\protect\\}

A $^{186}$WO$_2$ target was irradiated for 2.03 h.  Comparison of the DICEBOX-population calculations with the experimental-depopulation data for $^{187}$W sets $E_{\rm crit} = 900.0$~keV.  This value is substantially higher than in RIPL where $E_{\rm crit} = 145.9$~keV and includes only three levels. Table~\ref{tab:187Wg} lists 40 levels below $E_{\rm crit} = 900.0$ keV, deexcited by 121 secondary $\gamma$ rays and populated by 16 primary $\gamma$ rays, with a range of spins from $1/2 \leq J \leq 15/2$.  The capture state has $J^{\pi} = 1/2^{+}$.  Multipolarities and mixing ratios are taken from ENSDF \cite{basunia:09} where available or assumed according to $\Delta J$ and $\Delta \pi $ selection rules.


As was the case for the other tungsten isotopes investigated in this study,  $\Gamma_{0}$ shows a strong dependence on PSF/LD.  The EGLO/BSFG models give $\Gamma_{0} = 0.058(3)$~eV, which compares well with the adopted value of $\langle \Gamma_{0} \rangle = 0.051(5)$~eV \cite{mughabghab:06}.  For the EGLO/CTF combination $\Gamma_{0} = 0.038(2)$~eV, BA/CTF gives $\Gamma_{0} = 0.083(6)$~eV, and BA/BSFG gives $\Gamma_{0} = 0.127(7)$~eV.

A total thermal-capture cross section $\sigma_{0} = 33.33(62)$~b was determined for the $^{186}$W(n,$\gamma$) reaction.  Figure~\ref{fig:sigma0Ec_w187} shows the stability of this value with increasing cut-off energy, where $\sigma_{0}$ is nearly insensitive to $E_{\rm crit}$ even when as few as three levels are included.  For three levels and $E_{\rm crit} = 200$~keV ,we get $\sigma_{0} = 34.7(32)$~b.  Adopting $E_{\rm crit} = 900$~keV, with 40 levels in the decay scheme, $\sigma_{0}$ barely changes although the uncertainty is reduced by a factor of five.  The overall uncertainty of 1.9~\% is dominated by a 1.7~\% uncertainty in the simulated cross section with all other errors contributing less than 1~\%.  In Table~\ref{tab:187Wr} we compare our result with other measurements in the literature and the value adopted by Mughabghab of $\sigma_{0} = 38.1(5)$~b \cite{mughabghab:06}.  That value was based on an older activation decay-scheme normalization.  The literature values in Table~\ref{tab:187Wr} have been corrected for the decay-scheme normalization from our activation measurement, described in Section~\ref{sec:level4.E}, where possible.

\begin{table}
\caption{\label{tab:187Wr} Summary of $^{186}$W(n,$\gamma$) $\sigma_{0}$ measurements.}
\begin{tabular}{cc}
\hline\hline
$\sigma_{0}$ [b] & Reference \\ \hline
  {\bf 33.33(62)} & {\bf This work (prompt)} \\
  34.2(70) & L.~Seren \cite{seren:47} \\
  34.1(27) & H.~Pomerance \cite{pomerance:52}  \\
  41.3, 51 & W.~S.~Lyon \cite{lyon:60} \\
  33       & J.~H.~Gillette \cite{gillette:66} \\
  37.8(12) & S.~J.~Friesenhahn \cite{friesenhahn:66} \\
  35.4(8)  & P.~P.~Damle \cite{damle:67} \\
  40.0(15) & C.~H.~Hogg \cite{hogg:70} \\
  33.6(16)\footnotemark[1] & G.~Gleason \cite{gleason:77,exfor} \\ 
  33.3(11)\footnotemark[1] & R.~E.~Heft \cite{heft:78} \\ 
  37.0(30) & V.~A.~Anufriev \cite{anufriev:81} \\
  38.5(8)  & K.~Knopf \cite{knopf:87} \\
  34.8(3)  & M.~R.~Beitins \cite{beitins:92} \\
  34.7(15)\footnotemark[1], 37.9(20)\footnotemark[1] & S.~I.~Kafala \cite{kafala:97}\\
  32.7(10)\footnotemark[1]  & N.~Marnada \cite{marnada:99} \\
  32.8(10)\footnotemark[2] & F.~De~Corte \cite{decorte:03-2} \\ 
  30.6(19)\footnotemark[1] & M.~Karadag \cite{karadag:04} \\ 
  33.4(11)\footnotemark[2] & L.~Szentmikl{\'o}si \cite{szentmiklosi:06} \\ 
  35.9(11) & V.~Bondarenko \cite{bondarenko:08} \\
  38.7(23) & M.~S.~Uddin \cite{uddin:08} \\
  28.9(18)\footnotemark[1] & N.~Van~Do \cite{vando:08} \\ 
  29.8(32)\footnotemark[1] & A.~El~Abd \cite{elabd:10} \\ 
  38.1(5)\footnotemark[3]& Atlas~\cite{mughabghab:06} \\
 \hline\hline
\end{tabular}
\footnotetext[1]{Revised using the decay-scheme normalization determined in this work, $P_{\gamma}(685.7~{\rm keV}) = 0.352(9)$, see Section~\ref{sec:level4.E}.}
\footnotetext[2]{Weighted average from Table~\ref{tab:8}.}
\footnotetext[3]{Based on earlier decay-scheme normalizations.}
\end{table}

Figure~\ref{w187:PD}(a) shows excellent agreement between modeled population and experimental depopulation data for all levels except the 364.22-keV level.  This level was reported in ENSDF to  be deexcited by 162.7- and 286.9-keV $\gamma$ rays \cite{basunia:09}.  The {\small DICEBOX}-simulated population is much larger than the experimentally observed depopulation of this level.  Since the experimental data for all other levels compares well with their modeled populations over a range of five orders of magnitude, it is evident that the statistical model is an accurate simulation tool for the $^{187}$W capture-$\gamma$ decay scheme and discrepancies with the experimental intensity suggest incomplete experimental level or transition data.  The $J^{\pi} = 9/2^{-}$ assignment is firmly established for this level \cite{basunia:09}, so new $\gamma$ rays depopulating the 364.22-keV level were sought.  In Fig.~\ref{w187:PD}(b) we show that including a $\sim14$-keV transition populating the 350.43-keV level considerably improves agreement between experiment and theory. An additional low-energy $\gamma$ ray at 19.6~keV depopulating the 350.43-keV level is also suggested based on the statistical-model calculation.  These newly proposed $\gamma$-ray transitions were also inferred from the coincidence data of Bondarenko \emph{et al}. \cite{bondarenko:08}.

In an earlier ENSDF evaluation of $^{187}$W \cite{firestone:91} two additional levels were reported at 493.41 and 551~keV that were removed in the latest evaluation \cite{basunia:09}.  We see tentative evidence for the 143.2-keV $\gamma$ ray depopulating the 493.41-keV level.  The statistical model simulations imply a $J^{\pi} = 9/2^{-}$ assignment for this state.  There is insufficient evidence to support a level at around 551~keV, although there is a strong transition at 551.6~keV in the prompt capture-$\gamma$ spectrum.  This transition is also present in the delayed $^{187}{\rm W} \rightarrow ^{187}{\rm Re} + \beta^{-}$ beta-decay spectrum and can be attributed to the decay of $^{187}$Re.  We propose an additional 135.1-keV $\gamma$ ray depopulating the 775.60-keV level from the observed spectrum and consistency with statistical-model predictions.  An additional low-energy transition at 16.20~keV, with likely $E1$ multipolarity, is proposed to depopulate the 613.38-keV level based on statistical-model calculations.  The statistical model has also been used to estimate the intensity of the known 380.0-keV transition depopulating the $13/2^{-}$ level at 710.78-keV.  A doublet centered on 380.22~keV is observed in our data and we have resolved the intensity of the known 380.0-keV component by determining the intensity limit consistent with model predictions for a transition decaying out of this high-spin state.

The statistical-model simulations were also used to test uncertain $J^{\pi}$ assignments for levels in $^{187}$W.  The majority of the tentative $J^{\pi}$ assignments, for energy levels beneath $E_{\rm crit}$, were found to be consistent with the current ENSDF assignments, and 19 $J^{\pi}$ assignments for $^{187}$W \cite{basunia:09} could be confirmed in our analysis (see Table~\ref{tab:187Wg}).  A recent investigation of the $J^{\pi}$ assignments in $^{187}$W using polarized deuterons incident upon a natural tungsten foil to measure the (d,p) reaction~\cite{bondarenko:08} compared the observed particle angular distribution with DWBA calculations and determined $J$ and $l$-transfer values utilizing the {\small CHUCK3} code \cite{kunz:CHUCK3}.  Our results are consistent with most of the $J^{\pi}$ assignments from (d,p) analysis except for an excited state at 884.13~keV.  The (d,p) analysis suggests a value of $J^{\pi} = 7/2^{+}$ for this state, but we find that $J^{\pi} = 5/2^{+}$ is in agreement with our (n,$\gamma$) data, as illustrated in the population-depopulation plots in Fig.~\ref{w187:PD_2}.  The 884.13-keV state decays by a 474.02-keV transition, an assumed $E2$ quadrupole, to the 1.38-$\mu$s isomer at 410.06~keV, implying a likely $J^{\pi}=9/2^{+}$ assignment for this bandhead.  Consequently, all other members of the rotational sequence built on this level will have spin values increased by one unit of angular momentum, as shown in Fig.~\ref{w187:PD_2}.  The previous $J^{\pi} = (11/2^{+})$ \cite{basunia:09} assignment for the 410.06-keV isomer was based on the systematics of neighboring odd-$A$ tungsten isotopes.  Since only a few DWBA fits have been published, it would be instructive to see how well DWBA calculations for the lower-spin sequence would compare with the (d,p) data, as the shapes of experimental angular distributions are often well described by more than one set calculations, especially where counting statistics may be poor.

\begin{table*}
\caption{\label{tab:8} Partial $\gamma$-ray cross sections [b] and $P_{\gamma}$ values corresponding to decay lines observed in $^{187}$Re following the $\beta^{-}$ decay of $^{187}$W from this work and Refs.~\cite{basunia:09, szentmiklosi:06, decorte:03-2}.}
\begin{tabular}{|c|c|c|c|c|c|c|c|c|}
\hline\hline
$E_{\gamma}$ [keV] & $\sigma_{\gamma}^{(P)}$ \footnotemark[1] & $\sigma_{\gamma}^{(D)}$ \footnotemark[1] & $P_{\gamma}$ \footnotemark[2] & $P_{\gamma}$ \footnotemark[3] & $\sigma_{\gamma}$ \footnotemark[4] & $\sigma_{0}$ \footnotemark[5] & $\sigma_{\gamma}$ \footnotemark[6] & $\sigma_{0}$ \footnotemark[7] \\
\hline


134.34(7) & 3.60(12) & 3.66(12) & 0.110(4) & 0.104(2) & 3.65(7) & 33.2(14) & 3.50(2) & 31.9(12) \\

479.47(5) & 9.55(16) & 9.65(22) & 0.289(9) & 0.266(4) & 9.29(14) & 32.1(11) & 9.19(9) & 31.7(10) \\

551.22(9) & 2.16(19) & 2.20(4) & 0.0661(17) & 0.0614(10) & 2.16(4) & 32.6(10) & 2.14(1) & 32.37(85) \\

617.96(6) & 3.12(11) & 2.54(5) & 0.0762(21) & 0.0757(12) & 2.66(5) & 35.0(11) & 2.68(1) & 35.18(98) \\

625.03(10) & 0.35(11) & 0.419(19) & 0.0126(6) & 0.0131(2) & 0.47(1) & 37.2(20) & - & - \\

685.74(5) & 11.85(21) & 11.74(20) & 0.352(9) & 0.332(5) & 11.78(21) & 33.5(10) & 11.48(6) & 32.60(84) \\

772.99(10) & 1.606(95) & 1.771(57) & 0.053(2) & 0.0502(8) & 1.75(3) & 33.0(13) & 1.74(1) & 32.8(12) \\
\hline\hline

\multicolumn{1}{|c|}{Average $\sigma_{0}$} & \multicolumn{2}{c|}{33.36(62) \footnotemark[8]} & \multicolumn{3}{c}{} & \multicolumn{1}{|c|}{33.4(11) \footnotemark[9]} & \multicolumn{1}{c}{} & \multicolumn{1}{|c|}{32.8(10) \footnotemark[9]}\\
\hline\hline

\end{tabular}

\footnotetext[1]{This work: ($P$) prompt spectrum; ($D$) delayed spectrum.}
\footnotetext[2]{Calculated using $\sigma_{\gamma}^{(D)}$, this work, assuming $\sigma_{0} = 33.33(62)$~b.}
\footnotetext[3]{From ENSDF \cite{basunia:09} based on decay-scheme normalization by Marnada \textit{et al}. \cite{marnada:99}.}
\footnotetext[4]{From Szentmikl{\'o}si \textit{et al}. \cite{szentmiklosi:06}.}
\footnotetext[5]{Calculated using $\sigma_{\gamma}$, Ref.~\cite{szentmiklosi:06}, and $P_{\gamma}$ from this work.}
\footnotetext[6]{From De Corte and Simonits \cite{decorte:03-2}.}
\footnotetext[7]{Calculated using $\sigma_{\gamma}$, Ref.~\cite{decorte:03-2} and $P_{\gamma}$ from this work.}
\footnotetext[8]{Determined in prompt measurement.}
\footnotetext[9]{Statistical uncertainty is from a weighted average of all values plus an average 2.9~\% systematic error from our decay-scheme normalization.}

\end{table*}

We did not observe the 45.8(3)~keV, presumed $E1$ transition \cite{basunia:09}, deexciting 410.06-keV 1.38-$\mu$s isomer, that was reported by Bondarenko {\it et al}.~\cite{bondarenko:08} on the basis of delayed coincidences with the 474.02-keV $\gamma$-ray deexciting the 884.13-keV level.  Bondarenko {\it et al}. also postulated a second, $\sim 59$-keV transition, based on delayed coincidences with $\gamma$ rays deexciting the 350.43-keV, $7/2^{-}$ level.  This transition is of the same energy as the strong tungsten $K_{\alpha_{1}}$ X rays that obscure it in the spectrum.  Bondarenko {\it et al}. speculated the existence of the 59-keV $\gamma$-ray as unlikely since it required an $M2$ multipolarity assuming an $11/2^{+}$ assignment for the 410.06-keV level.  Our new $J^{\pi} = 9/2^{+}$ assignment for the 410.06-keV level implies an acceptable $E1$ transition for this 59-keV $\gamma$ ray.  However, the existence of the 59-keV $\gamma$-ray still remains in doubt since the proposed 13.80-keV transition deexciting the 364.22-keV level would also explain the coincidence results.  We observed two $\gamma$-rays populating the 410.06-keV isomer from higher levels below $E_{\rm crit}$.  The experimental intensity feeding the isomer, $\sum \sigma_{\gamma}^{\rm exp}(9/2^{+};410.06~{\rm keV}) = 0.394(16)$~b, together with the {\small DICEBOX}-modeled contribution from the quasi continuum, $P(9/2^{+};410.06~{\rm keV}) = 0.0145(14)$, yields a radiative thermal-capture cross section for the isomer $\sigma_{9/2^{+}} = 0.400(16)$~b.  This lower limit is consistent with our simulated population for $J^{\pi} =9/2^{+}$ and inconsistent with $J^{\pi} = 11/2^{+}$ (Fig.~\ref{w187:PD_2}).  Based on our analysis we propose new $J^{\pi}$ assignments for the five levels at: 410.06~keV ($9/2^{+}$); 493.4~keV ($9/2^{-}$); 597.24~keV ($11/2^{+}$); 815.51~keV ($13/2^{+}$); and, 884.13~keV ($5/2^{+}$).

\subsection{\label{sec:level4.E}Activation cross sections for $^{187}$W ($T_{1/2} = 24.000(4)$~h) \protect\\}

The same $^{186}$W target used in the prompt $\gamma$-ray measurements was later analyzed, offline, to determine the activation cross sections, $\sigma_{\gamma}$, for $\gamma$ rays emitted following $^{187}$W decay.  Since this measurement was performed in the same experiment, the decay $\gamma$-ray cross sections could be determined proportionally to the cross sections of the prompt $\gamma$ rays.  These activation $\gamma$-ray cross sections, together with their $\gamma$-decay emission probabilities, $P_\gamma$, independently determine the total radiative neutron-capture cross section, $\sigma_{0}$.

The decay $\gamma$ rays were observed in both the prompt spectrum, where the background from prompt $\gamma$ rays was high, and after bombardment, when the background was much lower. To determine the activation $\gamma$-ray cross sections, they must be corrected for saturation during bombardment, decay following bombardment and before counting begins, and decay during the counting interval.  The decay $\gamma$ rays, measured in the prompt spectrum, can be corrected with an in-beam saturation factor ($B$) defined as
\begin{equation}
B = 1 - \left( \frac{1-{\rm exp}(-\lambda t_{S})}{\lambda t_{S}} \right),
\label{pg2}
\end{equation}
where $\lambda=\ln(2)/T_{1/2}$ is the decay constant and $t_{S}$ is the irradiation period.  This expression is valid assuming a constant neutron flux.  Monitoring showed little power variation at the Budapest Research Reactor~\cite{belgya:pc} during our measurements.  The corrected activation $\gamma$-ray cross sections, measured in the prompt spectrum, are then given by 
\begin{equation}
\sigma_{\gamma}^{(P)} = \frac{\sigma_{\gamma}}{B}, 
\label{pg3}
\end{equation}
where $\sigma_{\gamma}$ is the uncorrected cross section observed during bombardment.

When the sample is analyzed offline the $\gamma$-ray cross sections in the delayed spectrum must also be corrected for saturation corresponding to in-beam exposure according to the factor $S = 1 - \exp(-\lambda t_{S})$.  The decay time $t_D$ following bombardment until analysis commences, introduces a further correction factor $D = \exp(-\lambda t_{D})$.  In addition, decay during the counting interval $t_C$  is corrected by a factor $C = [1 - \exp(-\lambda t_{C})]/(\lambda t_{C})$.  The overall correction factor accounting for saturation, decay, and counting intervals can then be applied to the cross sections of the decay $\gamma$-rays observed in the delayed spectrum as
\begin{equation}
\sigma_{\gamma}^{(D)} = \frac{\sigma_{\gamma}}{S \cdot D \cdot C}.
\label{pg7}
\end{equation}

In this work the irradiation time was $t_{S} = 7536$~s, and the source decayed for a time $t_{D} = 64859$~s before being counted for $t_{C} = 11645$ s. The activation $\gamma$-ray cross sections for the most intense transitions in the prompt and delayed spectra are shown in Table~\ref{tab:8}.  The prompt and delayed $\gamma$-ray cross sections were consistent.  We can then determine the $\gamma$-ray emission probabilities, $P_{\gamma} = \sigma_{\gamma} / \sigma_{0}$, using $\sigma_{0} =33.33(62)$~b from our prompt $\gamma$-ray measurement.  These probabilities are also listed in Table~\ref{tab:8} and are consistent with the $P_{\gamma}$ values from ENSDF \cite{basunia:09}, based on the decay scheme normalization of Marnada \textit{et al}. \cite{marnada:99}.  Using the $P_{\gamma}$ values from our activation data, we can then find independent total radiative thermal neutron-capture cross sections, $\sigma_{0} = \sigma_{\gamma} / P_{\gamma}$, based on the delayed-transition cross sections reported in the activation measurements of Szentmikl{\'o}si \emph{et al}. \cite{szentmiklosi:06} and De~Corte and Simonits \cite{decorte:03-2}.  In this approach, we find that our prompt measurement, $\sigma_{0} = 33.33(62)$~b, compares well with the weighted average of Szentmikl{\'o}si \emph{et al}. \cite{szentmiklosi:06}, $\sigma_{0} = 33.4(11)$~b, and also, with that of De~Corte and Simonits \cite{decorte:03-2}, $\sigma_{0} = 32.8(10)$.

\section{\label{sec:level7}Neutron separation energies\protect\\}

A byproduct of our analysis is the determination of neutron separation energies, $S_{\rm n}$, for $^{183,184,185,187}$W from the (n,$\gamma$) primary $\gamma$-ray energy measurements and the final-level energies taken from ENSDF.  These results, corrected for recoil, are shown in Table~\ref{tab:Sn} where they are compared with the recommended values of Wang \textit{et al}. \cite{Wang:AME}.  We present more precise determinations of $S_{\rm n}$ for $^{184,185}$W.

\begin{table}
\caption{\label{tab:Sn} Neutron-separation energies determined from (n,$\gamma$) reactions: $S_{\rm n} = E_{\gamma} + E_{f} + E_{r}$, where $E_{f}$ is the energy of the final level and $E_{r} = E_{\gamma}^{2}/2A$ is the recoil energy.  The weighted average for each nuclide is compared to the adopted value of Wang \textit{et al}. \cite{Wang:AME}.}
\begin{tabular}{cccc}
\hline\hline
Nuclide & $E_{\gamma}$ & $E_{f}$ & $S_{\rm n}$ \\
\hline
$^{183}$W & 6190.78(6) & 0.0        & 6190.88(6) \\
         & 6144.28(6) & 46.4839(4) & 6190.87(6)\\
         &            & Average    & 6190.88(6) \\
         &            & Adopted    & 6190.81(5) \\
\hline
$^{184}$W & 7410.99(14)   & 0.0          & 7411.14(14) \\
         & 7299.69(16)   & 111.2174(4)  & 7411.03(16) \\
         & 6507.63(16)   & 903.307(9)   & 7411.05(16) \\
         & 6408.60(12)   & 1002.49(4)   & 7411.20(13) \\
         & 6289.51(13)   & 1121.440(14) & 7411.06(13) \\
         &               & Average      & 7411.11(13) \\
         &               & Adopted      & 7411.66(25)\\
\hline
$^{185}$W & 5753.65(5)    & 0.0          & 5753.74(5)  \\
         &               & Adopted      & 5753.71(30)\\
\hline
$^{187}$W & 5466.47(12)   & 0.0          & 5466.55(12) \\
         & 5320.65(8)    & 145.848(9)   & 5466.57(8)  \\
         & 5261.67(9)    & 204.902(9)   & 5466.65(9)  \\
         & 4684.31(7)    & 782.290(19)  & 5466.66(7)  \\
         & 4662.94(27)   & 803.369(22)  & 5466.37(27) \\
         & 4650.27(8)    & 816.256(19)  & 5466.58(8)  \\
         & 4626.40(7)    & 840.205(16)  & 5466.66(7)  \\
         & 4574.67(7)    & 891.93(4)    & 5466.66(8)  \\
         &               & Average      & 5466.62(7)  \\
         &               & Adopted      & 5466.79(5)  \\
\hline\hline
\end{tabular}
\end{table}

\section{\label{sec:level6}Summary\protect\\}

The total radiative thermal neutron-capture $\gamma$-ray cross sections, $\sigma_{0}$, for the four major tungsten isotopes are summarized in Table~\ref{tab:summary}.  The cutoff energies, $E_{\rm crit}$, partial $\gamma$-ray cross sections, $\sum \sigma^{\rm exp}_\gamma$, simulated continuum GS feedings, $P({\rm GS})$, and simulated cross sections, $\sum \sigma^{\rm sim}_\gamma$, and an error budget are also given in Table~\ref{tab:summary}.  Our new cutoff energies exceed the RIPL-suggested $E_{\rm crit}$ values \cite{capote:09} for all isotopes.  These analyses have established that $\sigma_{0}$ is nearly independent of the assumed value of $E_{\rm crit}$, which is consistent with our earlier results for the palladium isotopes \cite{krticka:08}.

\begin{table*}
\caption{\label{tab:summary} Total radiative thermal neutron-capture cross sections, $\sigma_{0}$, for $^{182,183,184,186}$W from this work are compared with the recommended values of Mughabghab \cite{mughabghab:06}.  The critical energies, $E_{\rm crit}$, were determined from our comparison of experimental data with DICEBOX simulations.  The terms $\sum \sigma^{\rm exp}_{\gamma}$ and $\sum \sigma^{\rm sim}_{\gamma}$ are the total experimental and simulated partial $\gamma$-ray cross sections directly feeding the ground state from levels below and above $E_{\rm crit}$, respectively.  The {\small DICEBOX}-modeled population, per neutron capture feeding the ground state from the quasi continuum, is $P({\rm GS})$.  The individual contributions to the overall error budget are: $\delta_{A}$, the statistical uncertainty from experiment; $\delta_{B}$, the systematic uncertainty from the $\gamma$-ray self-attenuation correction; $\delta_{C}$, the systematic uncertainty from the normalization of the experimental cross sections; and $\delta_{D}$; the systematic uncertainty from Porter-Thomas fluctuations in the {\small DICEBOX} simulations.  The error $\delta_{D}$ includes the correlations between the uncertainties in $\sum \sigma_{\gamma}^{\rm exp}$ and $P({\rm GS})$ (see Equation~\ref{eq:3}).  The errors $\delta_{A}$, $\delta_{B}$, and $\delta_{C}$, were combined in quadrature to give the overall uncertainty on $\sum \sigma_{\gamma}^{\rm exp}$.}
\resizebox{\textwidth}{!}{
\begin{tabular}{|c|c|c|c|c|c|c|c|c|c|c|c|c|c|c|}
\hline\hline
\multicolumn{6}{|c|}{Cross section results this work} & Ref.~\cite{mughabghab:06}&\multicolumn{8}{c|}{Error Budget}\\
\hline
Target & $E_{\rm crit}$ [keV] & $\sum \sigma_{\gamma}^{\rm exp}$ [b] & $P({\rm GS})$  & $\sum \sigma_{\gamma}^{\rm sim}$ [b] \footnotemark[1] & $\sigma_{0}$ [b] \footnotemark[2] & $\sigma_{0}$ [b] \footnotemark[3] & \multicolumn{2}{c|}{$\delta_{A}$ [b]} & \multicolumn{2}{c|}{$\delta_{B}$ [b]} & \multicolumn{2}{c|}{$\delta_{C}$ [b]} & \multicolumn{2}{c|}{$\delta_{D}$ [b]}\\
\hline
$^{182}$W & 490.0 & 14.84(86) & 0.274(29) & 5.61(89) &
20.5(14) & 19.9(3)  & 0.42 & 2.07~\% & 0.66 & 3.24~\% & 0.34 & 1.66~\% & 0.89 & 4.34~\%\\
$^{183}$W & 1370.0 & 7.60(17) & 0.189(27) & 1.77(32) &
9.37(38) & 10.4(2)& 0.16 & 1.70~\% & 0.004 & 0.04~\% & 0.055 & 0.59~\% & 0.32 & 3.40~\% \\
$^{184}$W & 392.0 & 1.07(6) & 0.252(33) & 0.36(7) &
1.43(10) & 1.7(1) & 0.045 & 3.18~\% &  0  &  0  & 0.035 & 2.43~\% & 0.07 & 4.67~\% \\
$^{186}$W & 900.0 & 28.42(25) & 0.147(14) & 4.90(56) &
33.33(62) & 38.1(5) & 0.21 & 0.63~\% & 0.026 & 0.08~\% & 0.12 & 0.37~\% & 0.56 & 1.67~\%\\
 \hline\hline
\end{tabular}}
\footnotetext[1]{$\sum \sigma_{\gamma}^{\rm sim} = \sum \sigma_{\gamma}^{\rm exp}\times P({\rm GS})/(1+P({\rm GS}))$.}
\footnotetext[2]{$\sigma_{0} = \sum \sigma^{\rm exp}_{\gamma} + \sum \sigma^{\rm sim}_{\gamma}$.}
\footnotetext[3]{Recommended values \cite{mughabghab:06}.}
\end{table*}

Several combinations of photon strength function and level density formalisms were compared to the experimental data.  Total radiative widths of the capture state were found to be very model dependent.  For the compound $^{183,184}$W capture states, we could best reproduce the mean-adopted width $\langle \Gamma_{0} \rangle$ \cite{mughabghab:06} with the EGLO/CTF model combination.  In the cases of $^{185,187}$W, $\Gamma_{0}$ was best reproduced assuming the EGLO/BSFG combination.  All combinations involving BA gave much poorer agreement with the adopted $\Gamma_{0}$.

This analysis proposes several changes to the decay schemes for the compound tungsten isotopes $^{183,184,185,187}$W.  For $^{183}$W, one new $\gamma$ ray below $E_{\rm crit}$ is proposed, based on statistical-model simulations, and a tentative $J^{\pi}$ assignment is confirmed. The 309.49-keV, 5.2(3)-s, $11/2^{+}$ isomer in $^{183}$W was populated with a cross section of 0.177(18) b.  For $^{184}$W, one new $\gamma$ ray was placed in the decay scheme, based on our experiments, an additional low-energy transition is proposed from simulations, and four tentative $J^{\pi}$ assignments are confirmed.  Our analysis also indicates that the capture state in $^{184}$W is consistent with the composition $J^{\pi}_{\rm CS} = 1^{-}(\gtrsim 80~\%)$, $J^{\pi}_{\rm CS} = 0^{-}(\lesssim 20~\%)$, which is also consistent with the \emph{Atlas of Neutron Resonances} \cite{mughabghab:06}.  We find $J^{\pi} = 1^{-}$ the most likely assignment for the bound resonance at $-26.58$~eV, implying a likely capture-state spin composition of $J^{\pi}_{\rm CS} = 0^{-}(7.4~\%) + 1^{-}(92.6~\%)$.  The 1285.00-keV, 8.33(18)-$\mu$s, $5^-$ isomer in $^{184}$W was populated with a cross section of 0.0246(55)~b.  In $^{185}$W two new low-energy $\gamma$-ray transitions are proposed based on simulations, and three previous tentative $J^{\pi}$ assignments have been validated.  The 197.38-keV, 1.67(3)-min, 11/2$^+$ isomer in $^{185}$W was populated with a cross section of 0.0062(16) b.  For $^{187}$W, 19 of the previous $J^{\pi}$ assignments are confirmed and new $J^{\pi}$ assignments are proposed for five levels, including a new $9/2^{+}$ bandhead assignment at 410.06~keV that was previously assigned $(11/2^{+})$.  In addition, we reintroduced the 493.4-keV level, from an earlier ENSDF evaluation \cite{firestone:91}, and a new $\gamma$ ray depopulating this level based on tentative evidence in the capture-$\gamma$ spectrum.  There is also tentative evidence for a new transition at around 135.1~keV, depopulating the 775.60-keV level.  Our $^{187}$W simulations support inclusion of four new low-energy $\gamma$ rays, three of which were previously inferred in the work of Bondarenko {\it et al}. \cite{bondarenko:08}.  The 410.06-keV, 1.38(7)-$\mu$s, $11/2^{+}$ isomer in $^{187}$W was populated with a cross section of $0.400(16)$~b.  An analysis of the $\beta^{-}$-delayed $\gamma$-ray spectrum provided an independent decay-scheme normalization based on a new set of $P_{\gamma}$ measurements that compare well to the ENSDF decay-scheme normalization \cite{basunia:09}, adopted from the earlier work of Marnada \textit{et al}. \cite{marnada:99}.  Independent values of $\sigma_{0}$, consistent with our prompt measurement, were then determined based on our activation-data decay-scheme normlaization, thus providing confirmation of our approach.

The decay-scheme improvements suggested in this work will be used to improve the ENSDF nuclear-structure evaluations \cite{ensdf}, that contribute to the RIPL nuclear-reaction database \cite{capote:09}.  The new thermal-capture (n,$\gamma$) data will be added to the EGAF database \cite{firestone:06}.  These new data will also be used to help produce a more extensive and complete thermal-capture $\gamma$-ray library for the ENDF \cite{chadwick:11} neutron-data library.  Additional measurements of capture $\gamma$-rays from the rare isotope $^{180}$W(n,$\gamma$) are in progress and will complete our knowledge of the tungsten isotopes and resolve discrepancies in the measured $\sigma_{0}$ for this nucleus.

\section*{Acknowledgments}
This work was performed under the auspices of the University of California, supported by the Director, Office of Science, Office of Basic Energy Sciences, of the U.~S. Department of Energy at the Lawrence Berkeley National Laboratory under Contract DE-AC02-05CH11231, and by the U.~S.~Department of Energy by the Lawrence Livermore National Laboratory under Contract DE-AC52-07NA27344.  The access to the Budapest PGAA facility was financially supported by the NAP VENEUS08 grant under Contract OMFB-00184/2006.  Additional support was received through the research plan MSM 002 162 0859 supplied by the Ministry of Education of the Czech Republic.  The operations staff at the Budapest Research Reactor are gratefully acknowledged.

\bibliography{w_prc_2col_r}

\end{document}